\documentclass[pdflatex,sn-mathphys-num]{sn-jnl}

\usepackage{graphicx}%
\usepackage{multirow}%
\usepackage{amsmath,amssymb,amsfonts}%
\usepackage{amsthm}%
\usepackage{mathrsfs}%
\usepackage[title]{appendix}%
\usepackage{xcolor}%
\usepackage{textcomp}%
\usepackage{manyfoot}%
\usepackage{booktabs}%
\usepackage{algorithm}%
\usepackage{algorithmicx}%
\usepackage{algpseudocode}%
\usepackage{listings}%

\usepackage{dcolumn}   
\usepackage{braket}
\usepackage{float}
\usepackage{bm}        
\usepackage{standalone}
\usepackage{filehook}
\usepackage{gincltex}
\usepackage{collectbox}
\usepackage{filemod-expmin}
\usepackage{cancel}
\usepackage{tikz}

\theoremstyle{plain}%
\theoremstyle{remark}%
\newtheorem{remark}{Remark}%

\theoremstyle{definition}%
\newtheorem{definition}{Definition}%

\usepackage{calrsfs}
\DeclareMathAlphabet{\pazocal}{OMS}{zplm}{m}{n}

\usepackage{graphicx}
\makeatletter
\newcommand*\bigcdot{\mathpalette\bigcdot@{.5}}
\newcommand*\bigcdot@[2]{\mathbin{\vcenter{\hbox{\scalebox{#2}{$\m@th#1\bullet$}}}}}
\makeatother

\begin{document}

\title[From colored gravity to electromagnetism]{From colored gravity to electromagnetism}


\author*[1,2]{Robert Monjo} 
\affil[1]{Department of Mathematics, Saint Louis University. Manresa Hall, Calle Max Aub, 5, E-28003 Madrid, Spain 
\email{robert.monjo@slu.edu / rmonjo@ucm.es}}
\affil[2]{Department of Algebra, Geometry and Topology, Complutense University of Madrid, Plaza de Ciencias 3, E-28040 Madrid, Spain}

\author[3]{Álvaro Rodríguez-Abella}

\affil[3]{UCLA Department of Electrical and Computer Engineering, Los Angeles, CA 90095, USA
\email{rodriguezabella@g.ucla.edu}}

\author[4]{Rutwig Campoamor-Stursberg}

\affil[4]{Instituto de Matemática Interdisciplinar (IMI), Complutense University of Madrid, Plaza de Ciencias 3, E-28040 Madrid, Spain 
\email{rutwig@ucm.es}}


\abstract{The gauge formalism in \textit{telepalallel gravity} provides an interesting viewpoint to describe interactions according to an anholonomic observer's tetrad basis. Without going into assessing the complete viability of quantization in an early stage, this paper explores classical gravity within the framework of a classical-to-quantum bridge between the SU$(1, 3)$ Yang--Mills gauge formalism and the gauge-like treatment of teleparallel gravity. Specifically, the perturbed spacetime algebra with Weitzenböck connection can be assimilated to a local complexification based on the SU$(1,3)$ Yang--Mills theory, what we call hypercolor or, simply, color. The formulation of the hypercolor dynamics is build by a translational gauge, as in the teleparallel gravities. In particular, this work analyses small perturbations of a metric decomposition related to the Wilson line and the Kaluza--Klein metric, but obtaining electrodynamics in four dimensions. The spacetime coordinates are now matrices that represent elements of the $\mathfrak{su}(1,3)$ algebra. To make compatible the formulation of a colored gravity with the Lorentz force and the Maxwell equations, it is enough to define every energy potential origin as 0 in the event horizon instead of the classic zero potential at infinity. Under the colored gravity framework, standard electromagnetism can be obtained as a particular abelian case. }

\keywords{Teleparallel gravity, Weitzenböck connection, Yang–Mills theory}


\maketitle

\tableofcontents

\newpage

\section{Introduction}
\label{sec:introduction}

\subsection{Motivation}
\label{sec:motivation}



Geometrization is an exploration to extend the success of General Relativity (GR), involving geodesics of pseudo-Riemannian metrics, to other aspects of the nature. The geometrization of classical physics was a central axis of Wheeler \cite{Wheeler1966, Wheeler1961}, who stated that ``classical physics can be described in terms of curved empty space, and nothing more''. Wheeler introduced the notion of \textit{geons}, gravitational wave packets confined to a compact region of spacetime and held together by the gravitational attraction of the (gravitational) field energy of the wave itself. Wheeler was intrigued by the possibility that \textit{geons} could affect the test particles much like a massive object. This idea aimed to point the way to unify electrodynamics and general relativity. Aligned with this, it was deliberate the title of the Einstein's work ``On the electrodynamics of moving bodies'', and his later attempt to unify gravity and electromagmetism, also referred to as \textit{absolute (tele)parallelism} or \textit{teleparallel gravity} \cite{Einstein1928, Sauer2005, Trukhanova2020, Hohmann2023}. Since he stated that ``the gravitational field can be found only through the discovery of a logically simple mathematical condition,'' Einstein tirelessly attempted the formulation of a unified field theory for more than 30 years \cite{vanDongen2010}, employing the same methods \textcolor{black}{(see for instance the papers compiled in \cite{Delphenich2013})}.

As is well known, GR produces \textit{gravitomagnetic} effects and Maxwell's equations similar to classical electromagnetism \cite{Capozziello2009, Bakopoulos2014, Chatterjee2017, Adamek2020}. The awesome similitude between the primarily-observed natural force (Newtonian gravity) and the later Coulomb electrostatics guided several physicists for centuries to explore an early unification. As unifications require, gravity is the last one missing to be satisfactorily quantified, in spite of the remarkable contributions of (special and general) Relativity to Quantum Electrodynamics, such as the reference-frame (Lorentz) transformations, the Dirac spinor, and the \textit{Teleparallel Gravities} \cite{Hayashi1979, Sahin2016, Schwartz2023}. The Kaluza--Klein theory \cite{Williams2020} has been proposed as a candidate for a purely classical extension of General Relativity to $(\mathbb{R}^{1,4}, \widetilde {g})$, with metric $\widetilde{g}$, generalizing $(\mathbb{R}^{1,3}, {g})$ as follows: 
\begin{equation} \label{eq:kaluza_klein} 
{\widetilde {g}}_{\mu \nu }\equiv g_{\mu \nu }+\phi ^{2}A_{\mu }A_{\nu },\qquad {\widetilde {g}}_{5\nu }\equiv {\widetilde {g}}_{\nu 5}\equiv \phi^{2}A_{\nu },\qquad {\widetilde {g}}_{55}\equiv \phi^{2},
\end{equation} 
where $A^\mu$ is the electromagnetic vector field, $\phi$ is a scalar field such that $\square \phi =\frac{1}{4}\phi^{3}F^{\alpha \beta }F_{\alpha \beta }$, $F_{\alpha \beta } := \partial _{\alpha }A_{\beta}-\partial_{\beta }A_{\alpha}$ and $\square$ is the d'Alembert operator. The main achievement of the theory was to deduce, in a natural way, the free-field Maxwell equations and the electromagnetic stress-energy tensor to curve the spacetime. However, the gradient of $\phi$ leads to a dominant fifth component of the velocity, in contradiction to experience. 

If the particle quotient $q/m$ between charge $q$ and mass $m$ is incorporated into the Kaluza--Klein metric, it can explain some quantum effects, such as the Aharanov--Bohm effect and the London equation (see, for instance, \cite{Ferrari1989}). This effect consists of the modification of the matter wave, $\exp(i k_\mu dx^\mu)$, of a charged particle according to the Lagrangian $\exp(\mathrm{i}q A_\mu dx^\mu)$, where $k_\mu$ is the wave number and $A_\mu$ is the external electric potential.

The extra dimension of Kaluza--Klein theory was assimilated to the Lie group U(1) used in the gauge formalism of electromagnetism \cite{Krasnov2018}. A more general SU($n$) gauge theory was proposed by Yang--Mills (1954), and successfully implemented in the case of  SU(2)$ \times U(1)$ for \textit{Electroweak Dynamics} and SU(3) for \textit{Quantum Chromodynamics}. Currently, it is widely used in \textit{coupled gravity} theories \cite{Anderson2006, Guo2020} and integrated within the \textit{double-copy} or the \textit{squared} gauge theories \cite{Bern2010a, Bern2010b, Borsten2015, Anastasiou2017, Borsten2020, Monteiro2021}. These are \textit{supergravities} based on the original idea of Kaluza--Klein, that is, $g \sim A \otimes A$, where $A$ is a gauge potential (geometrically, a connection on either a vector bundle or a principal bundle), now linked to a color-kinematic duality. For instance, the \textit{Kerr--Schild--Kunndt metrics} contain a classic double copy of a Yang--Mills gauge field, which leads to a Einstein--Yang--Mills system \cite{Bern2010a, Gurses2018}.

In this context, M-Theory was for a certain time the most promising theory, but the extra dimensions (not yet observed) and the few proposed experiments make it difficult to test. Nevertheless, it uses an interesting property of the spacetime Lagrangian density for D-branes, firstly proposed by Born--Infeld for electrodynamics, 
\begin{equation}
\mathcal{L} \sim \alpha^{-2} \left(\sqrt{-\det\left(g_{\mu\nu} + \alpha F_{\mu\nu}\right)} \, - \, \sqrt{-\det(g_{\mu\nu})} \right) \;\; \xrightarrow{\alpha \to 0}\;\;  \frac{1}{4}  F_{\mu\nu}  F^{\mu\nu},
\end{equation} where $F_{\mu\nu}$ is the Faraday tensor or another field-strength curvature 2-form, while $\alpha$ is related to the tension of a string, D-brane or another quantity such as a \textit{dilaton coupling factor} \cite{Tseytlin2000, Kogan2003}. Moreover, this idea was used in Cosmology within the known as Eddington-inspired Born--Infeld gravity, which uses the symmetric part of the Ricci tensor, $R_{\mu\nu}$, instead of the field-strength, $F_{\mu\nu}$ \cite{Tamang2015}.

In contrast to String theories, Loop Quantum Gravity (LQG) uses only 4-dimensional manifolds $M$, however, taking into account the Wick-rotated (i.e. complex) spacetime. The LQG field states are invariant under the kinematical symmetry group $\mathrm{SU(2)} \times \mathrm{Diff}(M)$ of connection dynamics, that is, the semidirect product of the local SU(2) gauge transformations and the diffeomorphisms of $M$ \cite{Ashtekar2014}.

\textcolor{black}{Maxwell-type behavior was also found from the geometric structure developed by Itin \cite{Itin2006}, who constructed a coframe-based family of linear connections with six additional degrees of freedom that include both the Levi-Civita and the Weitzenböck connections as special limiting cases. For its dynamical propagation, it is necessary to satisfy a system of two first-order differential equations that coincides to the vacuum Maxwell system on a flat manifold.}

As a guiding principle of most theories, coupled Einstein--Maxwell equations are based on the assumption that the electromagnetic energy modifies the spacetime in accordance with the \textit{equivalence principle} or following the geometrization of the interactions \cite{Aldrovandi2006, Fuzfa2015}. GR states that the gravitational mass (energy) exactly equals in magnitude the inertial mass (energy); however, this may not be the case for all of the interactions. 

In this paper, we assume as a principle that \textit{all energy types contribute to a perturbation of the metric}, hence reinforcing that \textit{all interactions are due to the metric}. Gravity has no stress-energy tensor, but locally it is possible to define a work (energy) given by the gravity potential, which is transformed into kinetic energy. Both depend on the reference system and therefore are important actors in GR. The key idea is to define an adequate energy origin to perturb the metric. As a consequence, the interaction energy (mass) may be different from the inertial energy (mass). \textcolor{black}{However, the present work does not aim to stablish a ``theory of everything'' but it explores the link between teleparallel gravity and the SU(1,3) gauge theory under the above ideas. This step is probably necessary to better understand further developments in the direction of quantum gravity.}

The following paragraphs are devoted to reviewing the necessary material and notation used in this work (especially referred to the spacetime algebra and the spin connection), as well as to recall some fundamental concepts of the Yang-Mills theory. Sections 2 and 3 are devoted to the development of perturbations in the spin connection and in the spacetime algebra, respectively, focusing also on the macroscopic effects to reproduce the Maxwell equations and the Lorentz force. The final section summarizes the results and conclusions obtained. Unless otherwise stated, we use natural units ($\mathrm{c} \equiv 1 \equiv \hbar$).

\subsection{SU(1, $p$) Yang--Mills field equations}

A complex hyperbolic space $\mathbb{CH}^p$, also noted as $\mathbb{H}_{\mathbb{C}}^p$, is a Kähler manifold \cite{Pozetti2014, DiazRamos2023}, characterized by three mutually compatible structures: a complex structure, a Riemannian structure, and a symplectic structure. An important feature of $\mathbb{CH}^p$ is that they are symmetric spaces associated with pseudo-unitary groups $\mathrm{SU}(1, p)$, employed in modern quantum field theories. For instance, let \textit{correlators} or Green's \textit{correlation functions} be defined as vacuum expectation values of time-ordered products of quantum field operators. Lambert \textit{et al.} \cite{Lambert2021, Lambert2022} showed that $\mathrm{SU}(1, 3)$ conformal symmetry exhibits a rich structure, compared to conventional correlators in Lorentzian conformal field theories, and that they solve the Ward-Takahashi identities for general $N$-point correlation functions (see, for instance, \cite{Mouland2021}). 

Furthermore, Lipstein and Orchard \cite{Lipstein2022} found a reduction of 6d quantum gravity to a 5d Lagrangian theory with $\mathrm{SU}(1, 3) \times \mathrm{U}(1)$ spacetime symmetry, linked to the Poincaré algebra \cite{Lambert2022}. This idea is related to the Huang \cite{Huang2006} proposal for the unification of the gravitational and SU$(1,3)$ gauge fields, under the framework of a complex connection.

As for the special unitary groups SU($n$) with $n=1+p$, gauge symmetries of SU($1, p$) lead to Yang--Mills theories \cite{Anderson2006, Lambert2021, Lambert2022}. A parallel can be drawn between chromodynamics, which uses SU(3) as a structure group, and the present work that uses SU$(1, 3)$ as a structure group. For this reason, \emph{color} will be used for gauge fields in SU$(1, 3)$. The Yang--Mills developments are based on a \textit{gauge potential field} $A\in\Omega^1(P,\mathfrak{su}(1,p))$ that, geometrically, corresponds to a connection 1-form on a given \textit{principal \textrm{\normalfont SU$(1,p)$}-bundle}, $P\to M$, with $(M,\mathbf{g})$ being a Lorentzian manifold. Equivalently, gauge theories can be formulated on vector bundles. Let $\{\lambda_I\}_{I=1}^{n^2-1}$ be the \textit{(hyper)color} generators, i.e., the generators of a representation of $\mathfrak{su}(1,p)$. The structure constants $f_{IJ}^K$ encode the Lie algebra structure of $\mathfrak{su}(1,p)$: $[ \lambda_I, \lambda_J ] =if_{IJ}^K\lambda_K$.

Gauge transformations are defined by introducing a phase $\varphi(x)=\lambda_I\varphi^I(x)$, $x\in M$ to any physical wave $\Psi(x)$, that is,
\begin{equation}
\Psi(x)\mapsto\hat\Psi(x) = \exp(\mathrm{i} \varphi(x))\Psi(x).
\end{equation}
The invariance of physical measures of $\Psi(x)$ leads to a gauge covariant derivative, $\nabla_{\mu}$, which satisfies the (gauge) transformation $\nabla_{\mu}\Psi(x) \mapsto\widehat{(\nabla_{\mu}\Psi)}(x) = \exp(\mathrm{i} \varphi(x)) \nabla_{\mu}\Psi(x)$. Assuming that $(M,\eta)$ is the (flat) Minkowski spacetime and writing locally $A(x)=A_\mu(x) dx^\mu$, $x\in M$, the gauge covariant derivative is given by the following matrix operator, \begin{equation}
\nabla_{\mu}= \mathbf{1} \partial_{\mu} - i q \lambda_{I}A_{\mu}^{I} = \mathbf{1} \partial_{\mu} - i q A_{\mu},
\end{equation} where $A_{\mu} = \lambda_{I} A_{\mu}^{I}$, $\mathbf{1}$ is the identity matrix (matching the size of the algebra generators) and $q$ is a coupling constant.
The transformation of the gauge potential is given by
\begin{equation}
A_\mu(x)\mapsto\hat A_\mu(x)=U(x) A_\mu(x) U(x)^\dagger-\mathrm{i} q^{-1}(\partial_\mu U(x))U(x)^\dagger,
\end{equation}
where $U(x)=\exp(\mathrm{i} \lambda_I\varphi^I(x))$ and the subscript $\dagger$ denotes the Hermitian adjoint. Up to first order, the previous expression reads\begin{equation}\label{transfA}
A_{\mu}(x) \mapsto\hat A_{\mu}(x) =  A_{\mu}(x) - i[\lambda_J,\lambda_K]{A}_\mu^J(x)\varphi^K(x) + q^{-1}\lambda_I \partial_\mu \varphi^I(x).
\end{equation}

The strength of the field is geometrically the curvature of the gauge field, that is, the 2-form $\mathcal{F}=dA+A\wedge A\in\Omega^2(P,\mathfrak{su}(1,p))$. This curvature is a basic form of the adjoint type, which enables us to regard it as a form in the base space $M$, with values on the adjoint bundle, $\tilde{\mathfrak{su}}(1,p)=(P\times\mathfrak {su}(1,p))/\mathrm{SU}(1,p)$. It is given locally by the commutation of the covariant derivative
\begin{equation}
\mathcal{F}_{\mu\nu} := \mathrm{i}q^{-1}[\nabla_{\mu},\nabla_{\nu}] := \lambda_{I}\mathcal{F}_{\mu\nu}^{I} = \lambda_I\left(\partial _{\mu}A_{\nu}^{I}-\partial _{\nu}A_{\mu}^{I}-q f_{JK}^I A_{\mu}^{J}A_{\nu}^{K}\right).
\end{equation} As the relation $[\nabla_{\alpha},\mathcal{F}_{\nu\rho}^{I}]=\nabla_{\alpha}\mathcal{F}_{\nu\rho}^{I}$ holds, the Bianchi identity (Eq.~\ref{eq:Bianchi0}) and the Jacobi identity (Eq.~\ref{eq:Jacobi0}) are equivalent. 
\begin{eqnarray} \label{eq:Bianchi0}
(\nabla_{\mu}\mathcal{F}_{\nu\rho})^{I}+(\nabla_{\rho}\mathcal{F}_{\mu\nu})^{I}+(\nabla_{\nu}\mathcal{F}_{\rho\mu})^{I}=0,
\\ \label{eq:Jacobi0}
[\nabla_{\mu},[\nabla_{\nu},\nabla_{\rho}]]+[\nabla_{\rho},[\nabla_{\mu},\nabla_{\nu }]]+[\nabla_{\nu},[\nabla_{\rho},\nabla_{\mu}]]=0.
\end{eqnarray} 

Defining the dual strength as $\tilde{\mathcal{F}}^{\mu \nu }={\frac {1}{2}}\varepsilon ^{\mu \nu \rho \sigma }\mathcal{F}_{\rho \sigma }$, where $\varepsilon^{\mu\nu\rho\sigma}$ is the Levi-Civita tensor, the Bianchi identity reads $\nabla_\rho\tilde{\mathcal{F}}^{\mu \nu }=0$.

The gauge potential field has the property of being self-interacting and the equations of motion are semi-linear. This means that one can manage this theory only by perturbing with small nonlinearities.

In a four-dimensional manifold, by denoting $g = \mathrm{det}(\mathbf{g})$, the Lagrangian density without sources reads \begin{equation}
{\mathcal {L}}_{\mathcal{F}}=-{\frac {1}{2}}\sqrt{-g}\,{Tr} (\mathcal{F}^{2})=-{\frac {1}{4}}\sqrt{-g}\,\tilde{\mathcal{F}}^{\mu \nu }{\mathcal{F}}_{\mu \nu }.
\end{equation}
Similarly, if we consider $N$ spinor sources $\Psi=\{\Psi_k\}_{k=1}^N$ of energy $\{m_k\}_{k=1}^N$, it incorporates an additional term and no null equations of motion
\begin{equation} \label{lagYM}
{\mathcal {L}}_{{\mathcal{F}+\Psi} }=\sqrt{-g} \left(-{\frac {1}{4}}\tilde{\mathcal{F}}^{\mu \nu }{\mathcal{F}}_{\mu \nu } + \bar \Psi^k\left(i \gamma^\mu \nabla_\mu - m_k \right) \Psi_k\right) 
\end{equation} where $\bar \Psi^k := \gamma^0\Psi_k{}^{\dagger}$ and $\{\gamma^\mu\}_\mu$ are the gamma matrices. The corresponding field equations are \begin{equation}
\nabla_\mu\tilde{\mathcal F}^{\mu\nu} = J^\nu,
\end{equation}where $J^\nu : = \lambda_I J_I^{\nu}$ are the currents, with $J_I^{\nu}=q\bar \Psi^k\gamma^{\nu} \lambda_I \Psi_k$, which is equivariant under gauge transformations and satisfies the continuity equation $\nabla_{\mu}J_{I}^{\mu} = 0$.

\subsection{Affine, Cartan and Lorentz connections}
\label{sec:affine}

Some preliminaries are required to recall the notation used in modern relativity, especially in \textit{Lorentz connections} and \textit{teleparallel gravities} \cite{Pereira2012, Izaurieta2019, Krssak2019, Nazavari2023}. Let $(M,\mathbf g)$ be a 4-dimensional Lorentzian manifold. \textbf{Greek index labels} are used to denote a (general) reference frame, that is, a coordinate system on $M$: $\{x^\mu\}_{\mu\in\{0,1,2,3\}}$. The associated partial derivatives, which form a basis of local sections of $TM$, are denoted by $\{\partial_\mu\}_\mu$, and the corresponding dual basis of the cotangent bundle is denoted by $\{dx^\mu\}_\mu$.

An observer is given by a tetrad frame, which is a (local) orthonormal parallelization of $M$: $\{e_a\}_{a \in \{0,1,2,3\}} = \{e_0,e_1,e_2,e_3\}$. Note that we use \textbf{Latin index labels} for the tetrad frame. The tetrad components define a vierbein matrix, $(e^\mu_{\ a})_{a,\mu\in\{0,1,2,3\}}$, where $e_{a} = e^{\mu}_{\ a}\partial_{\mu }$ for $a\in\{0,1,2,3\}$. Analogously, the dual basis is known as the co-tetrad frame, $\{e^a = e_{\mu}^{\ a}dx^{\mu }\}_{a}$, with the dual relationship given by $e^{a}(e_{b})=e_{\mu}^{\ a}e^{\mu}_{\ b}=\delta _{b}^{a}$. The Lorentzian and Minkowski metrics are denoted by $\mathbf {g} = g_{\mu \nu } dx^{\mu }  dx^{\nu }$ and $\eta=\eta_{\mu\nu}dx^\mu dx^\nu\equiv diag(1,-1,-1,-1)$, respectively. Hence, the changes of the indexes are given by $e_{\mu}^{\ a} = g_{\mu\nu}\eta^{ab} e^{\nu}_{\ b}$ for $a,\mu\in\{0,1,2,3\}$, since $\eta_{ab}=\mathbf {g} \left(e_{a},e_{b}\right)=e_a\cdot e_b$ by definition \cite{Trukhanova2020}. 

\begin{remark}[Anholonomic observer frame]
A natural question is whether the tetrad frame can be regarded as the basis of partial derivatives associated to a certain reference frame. That is to say, if there exists a coordinate chart $\{x^a\}_{a\in\{0,1,2,3\}}$ such that $e_a=\partial_a$ for all $a\in\{0,1,2,3\}$ (whenever both vectors are defined). This can only be achieved if the Lorentzian manifold $(M,\mathbf g)$ is flat. In such a case, the frame $\{x^a\}_a$ is called an anholonomic observer, whereas the general frame $\{x^\mu\}_\mu$ is called a holonomic observer, and the vierbein matrix is given by $(e^{\mu}_{\ a}=\partial {x}^\mu / \partial {x}^a)_{a,\mu\in\{0,1,2,3\}}$. 
\end{remark}

The commutation relations of the derivatives in arbitrary coordinates are zero, $[\partial_\mu, \partial_\nu] = 0$, but tetrads $e_a := e^{\mu}_{\ a}\partial_\mu$ may present non-zero commutation relations: \begin{equation}
[e_a, e_b] = e^{\mu}_{\ a}e^{\nu}_{\ b} (\partial_\mu e_{\nu}^{\ c} - \partial_\nu e_{\mu}^{\ c}) e_c := \mathfrak{f}^c_{\ ab}e_c, 
\end{equation} where the $\mathfrak{f}^c_{\ ab}$ are known as the anholonomic coefficients. If dynamical constraints are not originated from coordinates (so called holonomic constraints), the system is called nonholonomic. A typical example of nonholonomic system is a disc rolling on a surface without slipping, since the new position of every specific point of the disc depends on the path taken, instead of the coordinates of the disc center.

The coordinate interval, which does not depend on the reference chosen, is given by $\mathbf{d}\boldsymbol{\tau} := \partial_\mu dx^{\mu} = e_a e^a\in TM\otimes T^*M$. Similarly, the invariant distance, that is, the proper time, $d\tau$, is found according to \begin{align*}  
d\tau^2 := \mathbf{d}\boldsymbol{\tau} \cdot \mathbf{d}\boldsymbol{\tau} & = (\partial_\mu\cdot\partial_\nu) dx^\mu dx^\nu =  g_{\mu \nu }dx^{\mu }dx^{\nu } = g_{\mu \nu }e^{\mu }_{\ a}e^{\nu }_{\ b} \, e^{a} e^{b}
\\ &= (e_a\cdot e_b)e^a e^b = \eta_{ab}e^{a} e^{b} = \eta_{ab} e_{\mu}^{\ a} e_{\nu}^{\ b} dx^{\mu}  dx^{\nu}. 
\end{align*} 
Observe that, in the previous expression, the metric is applied only to the first factor of the tensor product.

\begin{definition}[Affine connection]
Let $M$ be a smooth manifold and $V\to M$ be a vector bundle. An affine connection (or covariant derivative) on $V\to M$ is a map $D : \mathfrak{X}(M) \times \Gamma(V) \to \Gamma(V)$ denoted as $D_u\sigma$ for $u \in \mathfrak X(M)$ and $\sigma \in \Gamma(V)$, where $\mathfrak X(M)$ is the family of vector fields on $M$ and $\Gamma(V)$ is the family of sections of $V\to M$, such that (a) it is tensorial over $M$, i.e., $D_{fu + v}\sigma = fD_{u}\sigma + D_v\sigma$, (b) it is linear over $V$, i.e., $D_{u}(\sigma + \tau) = D_u\sigma + D_u\tau$, and (c) it satifies the Leibniz rule, i.e., $D_u(f\sigma) = u(f)\,\sigma + fD_u\sigma$, for all $f\in C^\infty(M)$, $u,v\in\mathfrak X(M)$ and $\sigma,\tau\in\Gamma(V)$.
\end{definition}

Note that an affine connection may also be regarded as a map $D:\Gamma(V)\to\Omega^1(M,V)$, where $\Omega^1(M,V)$ denotes the family of $V$-valued 1-forms on $M$. In a particular case, the Cartan connection is a principal connection on the frame bundle $FM \to M$ that is determined by an affine connection on the tangent bundle, $TM\to M$. In a reference frame, this affine connection can be written as $D = d + \Gamma$, where $d$ is the exterior derivative and $\Gamma:M\to T^*M\otimes End(TM)$ is a matrix of 1-form connection, where $End(TM)$ denotes the family of endomorphisms of $TM$. If we write $\Gamma=\Gamma_{\ \mu\nu}^\rho\,dx^\mu dx^\nu\partial_\rho$, where $\Gamma_{\ \mu\nu}^\rho$ are the Christoffel symbols of the connection on $TM\to M$, it can be decomposed as \begin{equation}\label{eq:decompositionconnection}
\Gamma_{\ \mu\nu}^\rho = \overset{\circ}{\Gamma}{}_{\ \mu\nu}^\rho + C_{\ \mu\nu}^\rho,    
\end{equation}
where $\overset{\circ}{\Gamma}{}_{\ \mu\nu}^\rho$ are the Christoffel symbols of the (torsion-free) Levi-Civita connection and $C_{\ \mu\nu}^\rho$ are the components of the so-called \textit{contorsion tensor}, which are given by \begin{equation}
C^{\rho}_{\ \mu\nu} = \frac{1}{2}\left(g^{\beta\rho}g_{\lambda\mu}T^{\lambda}_{\ \beta\nu}+g^{\beta\rho}g_{\lambda\nu}T^{\lambda}_{\ \beta\mu}-T^{\rho}_{\ \mu\nu} \right).
\end{equation} In the expression above, $T^{\rho}_{\ \mu\nu}$ are the components of the torsion tensor, which are given by $T^{\rho}_{\ \mu\nu} := \Gamma^{\rho}_{\ \mu\nu} - \Gamma^{\rho}_{\; \nu\mu}$.

Since the frame bundle is a principal $\textrm{GL}(4)$-bundle, the Cartan connection is (locally) determined by a 1-form $\omega\in\Omega^1(M,\mathfrak{gl}(4))$. If we restrict $\mathfrak{gl}(4)$ to $\mathfrak{so}(1, 3)$, the Lie algebra of the Lorentz group, $\textrm{SO}(1, 3)$, the corresponding principal connection is known as the Lorentz connection. Geometrically, we are considering a reduction of the principal $\textrm{GL}(4)$-bundle, $FM\to M$, to a principal $\textrm{SO(1, 3)}$-bundle, $P\to M$, which exists whenever $M$ is orientable.

Let $\{S_{ab}\}_{a,b\in\{0,1,2,3\}}$ be the generators of a spin representation of $\mathfrak{so}(1, 3)$ onto $\mathbb C^4$. In this case, we can write $\omega=\omega_\mu dx^\mu$ with $\omega_{\mu} = \frac{1}{2}\omega_{\mu }^{\ ab} S_{ab}$ for some functions $\omega_\mu^{\ ab}$ that are skew-symmetric in the algebraic indices \cite{Pereira2012}. Analogously, the components of the curvature and torsion 2-forms are given by $R_{\mu\nu} = \frac{1}{2} R^{ab}_{\ \ \mu\nu}S_{ab}$ and $T_{\mu\nu} = T^a_{\ \mu\nu} e_a$, respectively. For the latter,  observe that the generators of the Lie algebra of the \textit{translation group} are $\{e_{a}\}_{a}$. The Lorentz connection (sometimes called spin connection) satisfies the metricity condition (or metric compatibility) for the Riemann-Cartan geometry, and it can be particularized to the (torsionless) \textit{Ricci coefficient of rotation} to obtain GR. As stated above, the spin-connection components are obtained from the affine connection on the tangent bundle. Specifically, they are given by\begin{equation}\label{connection} 
\omega_{\mu }^{\ ab}=\eta^{bc}e_{\nu }^{\ a}(\partial _{\mu }   e_{\ c}^{\nu}+{\Gamma}{}_{\ \rho \mu }^{\nu }  e_{\ c}^{\rho}).
\end{equation} This connection defines the Fock--Ivanenko covariant derivative on the associated vector bundle $(P\times\mathfrak{so}(1, 3))/\textrm{SO}(1, 3)\to M$, \begin{equation} \label{eq:covariant}
D_{\mu} = \partial_{\mu} - {\textstyle \frac{\mathrm{i}}{2}} \omega_{\mu}^{\ ab} S_{ab}.
\end{equation} For a scalar field, the generators are given by $S_{ab} = 0$. Likewise, for a Dirac spinor field, we have $S_{ab} = {\frac{\mathrm{i}}{4}} [\gamma_a, \gamma_b]$. Finally, for a vector field, they are given by $(S_{ab})^c_{\ d} := \mathrm{i}(\eta_{bd}\delta^c_a - \eta_{ad} \delta^c_b)$. In particular, the Fock-Ivanenko covariant derivative of a vector field $(w^a)$ reads
\begin{equation} \label{eq:covariant2}
D_{\mu}w^a = \partial_{\mu}w^a + \eta_{bc}\omega_{\mu}^{\ ab} w^c.
\end{equation}In any case, the curvature of the Fock-Ivanenko covariant derivative is given by
\begin{equation} \label{curvature}
R_{\ \ \mu \nu}^{ab} := [D_\mu, D_\nu]^{a,b} = \partial_\mu \omega_{\nu}^{\ ab} - \partial_\nu \omega_{\mu}^{\ ab} + [\omega_{\mu},\omega_{\nu}]^{a b} \equiv (d\omega + \omega \wedge \omega)_{\mu \nu}^{\ ab},
\end{equation}where $[\omega_{\mu},\omega_{\nu}]^{a b} = \omega_{\mu}^{\ a}{}_{c}\omega_{\nu}^{\ cb} - \omega_{\nu}^{\ a}{}_{c}\omega_{\mu}^{\ cb} $. The Ricci scalar curvature is given by $R = R_{\ \ \mu \nu}^{ab} e^{\mu}_{\ a}e^{\nu}_{\ b}$.

\vspace{10mm}

\subsection{Teleparallel gauge-like theories}


Teleparallel gravities are a large set of theories that generalize or modify the GR \cite{Krssak2019, Bhatti2023}. The most widely studied is the \textit{teleparallel equivalent of General Relativity} (TEGR), which assumes that gravity is completely modeled by torsion instead of curvature. That is to say that the Riemann curvature is zero ($R^{ab}_{\ \ \mu\nu} = 0$) with null connection ($\omega_\mu^{\ ab}= 0$). TEGR Euler-Lagrange equations can be obtained from its action, which is dynamically equivalent to the Einstein-Hilbert equation, as they differ only by a boundary term \cite{Krssak2019, Andrade2000, guzman2024}.

Imitating gauge theories, some authors use translational invariance properties to formulate the TEGR theory. However, its ``translation-gauge formalism'' does not gather the commonly accepted requirements for a complete gauge theory \cite{Fontanini2019}. Independently of the mathematical interpretation, we employ this \textit{gauge-like} formalism due to its very interesting similarities with the classical gauge theories of particle physics. 

As in the previous sections, let $(M,\mathbf{g})$ be a Lorentzian manifold and assume that to each point of $M$ there is an attached Minkowski space. The anholonomic coordinates of the Minkowski space are denoted by $\{x^a\}_a$. The result is known as the Minkowski bundle $\mathscr M\to M$ with coordinates $\{x^\mu,x^a\}_{\mu,a}$. The translation group acts locally on the Minkowski bundle $x^a \mapsto \hat x^a=x^a + \epsilon^a(x^\mu)$. The Lie algebra of the translation group is generated by the partial derivatives of the anholonomic coordinates, $\{\partial_a\}_a$. The fields of teleparallel gravity are the 1-forms on $M$ with values on the Lie algebra of the translation group, which are referred to as \textit{translational gauge-like potentials}, denoted by $\phi=\phi_{\mu} dx^\mu$ with $\phi_\mu=\phi_{\mu}^{\ a} {\partial}_a$. An infinitesimal translation, $\delta x^a=\epsilon^a(x^\mu)$, yields a transformation of the gauge-like potential such that ${\hat\phi}_{\mu}^{\ a} = \phi_{\mu}^{\ a} - \partial_\mu \epsilon^a$. That is to say, the action of the translation group is extended to the family of gauge-like potentials. In addition, the gauge-like potential $\phi_\mu=\phi_\mu^{\ a}\partial_a$ induces a transformation of the vierbein matrix as follows (see e.g. \cite{Aldrovandi2006}): 
\begin{eqnarray} \label{eq:tetrad}
e_{\mu}^{\ a} = \partial_{\mu} x^{a}  \hspace{4mm} & \mapsto & \hspace{4mm}  \hat e_{\mu}^{\ a} = e_{\mu}^{\ a} + \phi_{\mu}^{\ a} 
    \hspace{3mm} = \hspace{3mm}
\partial_{\mu} x^{a} + \phi_{\mu}^{\ a}.
\end{eqnarray} 
This enables us to define the gauge-like derivative in TEGR as $D_{\mu} = \hat e_{\mu}^{\ a}\partial_{a}=\partial_\mu+\phi_\mu$. The corresponding strength field is given by $F_{\mu\nu} = F_{\ \mu\nu}^{a} \partial_a$, where $F^a_{\ \mu\nu}=\partial_\mu\phi^a_{\ \nu}-\partial_\nu\phi^a_{\ \mu}$.

On the other hand, the Weitzenb\"ock connection is the affine connection on $TM\to M$ defined from the new tetrad as $\Gamma^{\rho}_{\ \mu\nu} = \hat e_{\ a}^{\rho} \partial_{\nu} \hat e^{\ a}_{\mu}$. The curvature of this connection vanishes, whereas its torsion is given by $T^{\rho}_{\ \mu\nu} := \Gamma^{\rho}_{\ \nu\mu} - \Gamma^{\rho}_{\ \mu\nu}$. It turns out that the torsion of the Weitzenb\"ock connection is the field strength of the translational gauge-like potential seen from the anholonomic frame:
\begin{equation} \label{eq:strength_transl}
F^a_{\ \mu\nu}=\hat e_\rho^{\ a} T^\rho_{\ \mu\nu}=(e_\rho^{\ a}+\phi_\rho^{\ a})T^\rho_{\ \mu\nu}
\end{equation}

TEGR assumes that the inertial effects are represented by the  Weitzenb\"ock connection, while the gravity force is given by the contorsion tensor. In fact, Eq. \ref{eq:decompositionconnection} yields a decomposition of the geodesic equation in General Relativity:\begin{equation}
0 = \frac{d u_\mu}{d\tau} - \overset{\circ}{\Gamma}{}^\rho_{\ \mu\nu} u_{\rho} u^{\nu} = \frac{d u_\mu}{d\tau} - \Gamma^\rho_{\ \mu\nu} u_{\rho} u^{\nu} + C^\rho_{\ \mu\nu} u_{\rho} u^{\nu},
\end{equation}
where $u^\mu=d x^\mu/d\tau$ are the components of the 4-velocity. As we see, the Levi-Civita connection contains both inertial effects and gravity forces under the GR framework. Likewise, the dynamical equations for a spinless particle of mass $m$ in a gravitational field $\phi_\mu^{\ a}$ are \cite{deAndradePereira1997}\begin{equation}
\hat e_{\mu}^{\ a} \frac{d u_a}{d\tau} \; = \;\frac{d u_\mu}{d\tau} - \Gamma^\rho_{\ \mu\nu} u_{\rho} u^{\nu} =  - C^\rho_{\ \mu\nu} u_{\rho} u^{\nu}  \; = \; T^\rho_{\ \mu\nu} u^{\rho} u^{\nu}  \; = \; F_{\ \mu\nu}^{a} u_{a} u^{\nu},
\end{equation}where $u^a=\hat e_\mu^{\ a}u^\mu$ are the anholonomic components of the 4-velocity of the particle. Finally, the TEGR Lagrangian density is given by \cite{deAndradePereira1997, Krssak2019},\begin{equation} \label{eq:lagrangian_TEGR}
\mathcal{L} = \frac{\hat e}{2\kappa}\left(\frac{1}{4}T^\rho_{\ \mu \nu} T_{\rho}^{\ \mu\nu} + \frac{1}{2}T^\rho_{\ \mu\nu}T^{\nu\mu}_{\ \ \rho} - T^\rho_{\ \mu\rho}T^{\nu\mu}_{\ \ \nu}\right) = \frac{\hat e}{2\kappa}\left(\frac{1}{4}F^a_{\ \mu\nu} F_{a}^{\ \mu \nu}\right),
\end{equation}
where $\kappa = 8\pi\mathrm{G}$ and $\hat e:=\det {\hat e_{\mu }}^{\ a} = \sqrt {-g}$. 
This Lagrangian density provides the same dynamics as the Einstein--Hilbert Lagrangian with Ricci scalar $R$, that is, ${\mathcal{L}_{GR}} = \frac{1}{\kappa} \hat e R $, since it only differs in a divergence, ${\mathcal{L}} - \mathcal{L}_{GR} = - \frac{1}{\kappa} \partial_{\mu}(\hat e\,T^{\nu\mu}_{\ \ \ \nu})$.

Observe that the TEGR Lagrangian density is very close to the Lagrangian density of a Yang--Mills theory with group structure $U(1)$. Notice that, from the equations above, it is easy to choose some gauge-like potential $\phi_\mu$ to imitate the electrodynamic gauge $A_\mu$ such as, for instance, $\phi_\mu^{\ a} \sim A_{\mu} u^a$ with an additional factor $\frac{q}{m}$ to connect the energy $m$ and the coupling factor $q$. This fact will be central in the following.

\vspace{10mm}

\subsection{Spacetime algebra}

We briefly summarize the main facts (see \cite{Gu2018, Gu2018b} for more details). Let $(M,\mathbf{g})$ be a 4-dimensional Lorentzian manifold and $\mathscr M\to \mathrm{T} M$ be the Minkowski bundle with frames $\{x^\mu,x^a\}_{\mu,a}$, as in the previous section. Recall that the spacetime algebra, also known as the Lorentz algebra, is the Clifford algebra $Cl_{1,3}(\mathbb{R},\mathbf{h})$, with $\mathbf{h}=\eta$ or $\mathbf{g}$. Let $\{{v}_\mu\}_\mu$ and $\{\gamma_a\}_a$ be the generators of a representation of the spacetime algebra in the Lorentzian and the Minkowski metrics, respectively. Hence, they satisfy the following relations
\begin{eqnarray}
{v}_{\mu } \bigcdot {v}_{\nu } := {\frac{1}{2}} ({v}_{\mu}{v}_{\nu}+{v}_{\nu } {v}_{\mu })=g_{\mu \nu } \mathbf{1}_4\\
\gamma_{a} \bigcdot \gamma_{b} := {\frac{1}{2}} ({\gamma}_{a}{\gamma}_{b}+{\gamma}_{b} {\gamma}_{a})=\eta_{ab} \mathbf{1}_4
\end{eqnarray}
where $\bigcdot$ is known as the outer product and $\mathbf{1}_4$ denotes the (sometimes omitted) identity matrix. Besides, the exterior product is defined as
\begin{eqnarray}{v}_{\mu } \wedge {v}_{\nu } := { \frac{1}{2}} ({v}_{\mu}{v}_{\nu}-{v}_{\nu } {v}_{\mu }) = { \frac{1}{2}} [{v}_{\mu }, {v}_{\nu }]\\
\gamma_{a} \wedge \gamma_{b} := { \frac{1}{2}} ({\gamma}_{a}{\gamma}_{b}-{\gamma}_{b} {\gamma}_{a}) = { \frac{1}{2}} [{\gamma}_{a}, {\gamma}_{b}]
\end{eqnarray}
Therefore, the decomposition into the symmetric and skew-symmetric parts is given by ${v}_{\mu} {v}_{\nu} = {v}_{\mu} \bigcdot {v}_{\nu} + {v}_{\mu} \wedge {v}_{\nu}$, and analogous for $\gamma_a\gamma_b$.  
The relationship between tetrad components and spacetime generators is given by ${v}^{\mu} = e^{\mu}_{\ a}  \gamma^a$ and $\gamma^{a} = e_{\mu}^{\ a}  {v}^{\mu}$. They are also related to the Lorentz or spin connection.  If we define $\gamma^{a} := \gamma_a^{-1}$, the orthogonality property is satisfied, i.e., $\gamma_{a}\bigcdot\gamma^{b} = \delta_{a}^b\mathbf{1}_4$, and the same holds for $v_\mu$. In particular, ${v}_{\mu}\bigcdot{v}^{\mu} = \gamma_a\bigcdot\gamma^a =\mathbf{1}_4$.

The symbol $v_\mu$ for the generators of the representation of $Cl_{1,3}(\mathbb R,\mathbf g)$ is purposely chosen because of its link to the four-velocity covector $u_\mu := dx_{\mu}/d\tau$ with the line element $d\tau := \sqrt{g_{\nu\mu}dx^{\nu} dx^{\mu}} = \sqrt{dx_{\mu} dx^{\mu}}$, whereas the symbol $\gamma_a$ is used in honour to the representation by the Dirac gamma matrices. Indeed, distances defined by the generators of the spacetime algebra are analogous to the ones defined by the co-vector $u = (u_{\mu})_{\mu \in \{0,1,2,3\}}$, that is, $d\tau = u_{\mu}dx^{\mu} \in \Omega^1 (M)$. Namely, the element of spacetime is now defined as $\mathbf{d\boldsymbol\tau} := {v}_{\mu} dx^{\mu} = \gamma_{a} d{x}^{a}$. This enables us to see the spacetime metric as the inner product of the coordinate tangent co-vector fields, where the inner product is defined from the outer product of the algebra generators. More specifically,
\begin{align*}
\mathbf{1}_4\,d\tau^2 = \mathbf{d\boldsymbol\tau} \bigcdot \mathbf{d\boldsymbol\tau} &= ({v}_{\mu }  \bigcdot {v}_{\nu }) dx^{\mu} dx^{\nu} = \mathbf{1}_4\,g_{\mu \nu} dx^{\mu} dx^{\nu} \\
&= ({\gamma}_{a}  \bigcdot {\gamma}_{b}) d{x}^{a} d{x}^{b}=\mathbf{1}_4\,\eta_{a b}  d{x}^{a} d{x}^{b}
\end{align*}

For each vector field $(w^a)$, the total time derivative is given by
\begin{equation} \label{totaltime}
\mathbf{D_{\tau}} w^a :=({v}^{\mu} D_{\mu})w^a = \mathbf{d_{\tau}}w^a + \eta_{bc}  \omega_{\mu}^{\ ab} w^c {v}^{\mu}
\end{equation} where $D_\mu$ is the Fock-Ivanenko covariant derivative, and $\mathbf{d_{\tau}} = {v}^{\mu}\partial_{\mu} = \gamma^a{\partial}_{a}$. Observe that the classical definition is obtained in straightforward manner by replacing $v^{\mu} \mapsto u^{\mu}$ and the outer product by the standard product.


The main hypothesis of this paper is that the Cartan covariant derivative for gravity behaves like an $\textrm{SU}(1,p)$ gauge derivative for $p = 3$. For this purpose, the spacetime generators $\gamma_a$ are perturbed by an $\textrm{SU}(1,p)$ gauge potential field $A_\mu$ that involves the gauge phase of matter-energy fields such that $v^\mu \approx e^\mu_{\ a}{\gamma}^a + O(A_\mu)$. In other words, we assume that the set of generators $\{{v}_{\mu}\}_{\mu}$ contains information about some dynamics of the Standard Model. Our goal is to show how the $\textrm{SU}(1,p)$ Yang--Mills dynamics emerges from locally-perturbed tetrads, and to check some related features, such as the recovery of electrodynamics.  

\subsection{Complexified Minkowskian spacetime}

Let $(\mathscr{M},\eta)$ be the Minkowskian manifold with metric $\eta = \mathrm{diag}(1, -1, -1, -1)$ and $(\mathscr{M}^c,\eta^c)$ be its \textit{complexification}, defined by $\mathscr{M}^c := \mathscr{M} \oplus \mathbf{i}\,\mathscr{M}$, where $\mathbf{i} := e_0 e_1 e_2 e_3$ is the unitary pseudoscalar obtained with the vector basis $\{e_\mu\}_{\mu=0}^4$ of $\mathscr{M}$ \cite{Friedman2021, Chappell2023}. In contrast to the \textit{real} Minkowskian four-vectors (e.g. velocity and linear momentum) that do not depend on the space orientation choice, the \textit{imaginary} part consists of four-pseudovectors (e.g. angular and magnetic momenta) whose direction is affected by the change of orientation (parity).The terminology of \textit{imaginary} and \textit{complexification} is due to the property of $\textbf{i}^2 = (e_0 e_1 e_2 e_3)(e_0 e_1 e_2 e_3) = -1$, with $e_ie_j := e_i \cdot e_j + e_i \wedge e_j$.

The inner product structure on $\mathscr{M}^c \ni \mathrm{a}, \mathrm{b}\;$ is defined by 
\begin{equation*}
    \eta(\mathrm{a}, \mathrm{b}) := \eta(a+\mathbf{i}\,x,\;  b + \mathbf{i}\,y) := \eta(a,b)+\eta(x,y)-\mathbf{i}\,\eta(x,b) + \mathbf{i}\,\eta(a,y)\,,
\end{equation*}
where $\mathrm{a} = a+\mathbf{i}\,x,\;  \mathrm{b} = b + \mathbf{i}\,y$ and $a,b,x,y$ are real tangent (vector) fields over $\mathscr{M}$. For instance, any spinor (spin-$\frac{1}{2}$ particle) can be described as $\psi = u + \mathbf{i}\,s\, \in \mathscr{M}^c$, where $u$ is the unitary \textit{linear momentum} or four-velocity of the particle and $s$ is the four-dimensional (spin) angular momentum, which is a Pauli-Lubanski pseudovector \cite{Chappell2023} satisfying $s^2 = -1$ and the orthogonality $u \cdot s = \eta(u,s)= 0$, while $u^2 = 1$. This relativistic two-state prescription is useful for describing, for example, the interaction of a $q$-charged particle with mass $m$, spin $s$ and Land\'e factor $\mathrm{g_e}\approx 2$. Let $F$ be an electromagnetic tensor; then the interaction with current $q\psi$ is $\frac{d}{d\tau}\psi^\alpha \approx \, \mu \, \psi^\beta F_{\beta}^{\ \alpha}$, where $\mu = \frac{\mathrm{g}_e}{2}\frac{q}{m}s \approx \frac{q}{m}s$ is the magnetic momentum \cite{Friedman2021}.

\section{Colored gravity}

\subsection{Complexified Lorentzian metric}
\label{sec:complex}

\noindent In the following, all the notations previously introduced are assumed. Moreover, let $(M, g)$ be an oriented Lorentzian 4-manifold of signature $(+, -, -, -)$ with a \textit{spin} or oriented loop space $\pazocal{L}M$.

Let $\mathrm{T}M$ be the real tangent bundle of $M$ and $\mathrm{T}^cM = \mathrm{T}M\otimes \mathbb{C} \cong \mathscr{M}^c$ be its complexification. \textcolor{black}{The corresponding frame bundle, which is a principal GL$(4, \mathbb{C})$-bundle, is denoted by $L^c(M)\to M$.} The real metric $g$ on $\mathrm{T}M$ has a unitary extension on $\mathrm{T}^cM$ given by 
\begin{equation*}
    g(a+\mathbf{i}\,x,\;  b + \mathbf{i}\,y) := g(a,b)+g(x,y)-\mathbf{i}\,g(x,b) + \mathbf{i}\,g(a,y)\,,
\end{equation*}
where $a,b,x,y$ are real tangent fields over $M$. The resulting metric preserves the signature of the original metric and thus it is Lorentzian of signature $(+,-,-,-)$. Therefore, the metric induces a reduction of \textcolor{black}{$L^c(M)\to M$ to a principal U$(1,3)$-bundle. Similarly, the volume form induced by $g$ and the orientation on $M$ yield a reduction of $L^c(M)\to M$ to a principal SL$(4, \mathbb{C})$-bundle. By gathering both, a reduction\footnote{\textcolor{black}{Recall from \cite[Proposition 5.6]{kobayashinomizu1963} that the existence of this reduction is equivalent to the existence of a cross section of the bundle of oriented orthonormal frames of $M$, i.e., the associated bundle 
\begin{equation*}
\frac{L^c(M)\times{\rm GL}(4,\mathbb C)/{\rm SU}(1,3)}{{\rm GL}(4,\mathbb C)}\to M,
\end{equation*}
where the (well-defined) action of ${\rm GL}(4,\mathbb C)$ on ${\rm GL}(4,\mathbb C)/{\rm GL}(1,3)$ is given by left multiplication.}} of $L^c(M)\to M$ to a principal bundle with structure group $\mathrm{U}(1,3) \cap \mathrm{SL}(4,\mathbb{C}) = \mathrm{SU}(1,3)$ is obtained. In Sect. \ref{sec:perturbedspinor}, we onsider a principal connection $A$ on this principal bundle.}

The Levi--Civita connection $\overset{\circ}D$ of $g$ extends as $\overset{\circ}D{}^c$ to $\mathrm{T}^cM$, thereby providing some reference (torsion-free) connection among all SU$(1, 3)$ connections on $\mathrm{T}^cM$. Under this framework, we develop a theory based on the non-compact gauge group SU$(1,3)$, which is subject to long-standing debate concerning the properties of a \textit{quantum field theory} (QFT) underlying such a gauge theory \cite{Tseytfin1995,weinberg2005,Fabbrichesi2021}.The non-compact nature of a group such as the U(1) gauge group is \textcolor{black}{usually} managed through the Faddeev-Popov technique with gauge fixing (e.g. Lorenz gauge) and ghost fields for preserving unitarity, which help to identify the physical degrees of freedom of the bosons and ensure consistency in the quantization process \cite{Thierry1980,Eichhorn2013}. Another solution for a quantum geometrodynamics is to slice the covariant spacetime into 3-dimensional spacelike hypersurfaces, and its 1-dimensional transverse temporal flux, such as in the Arnowitt–Deser–Misner (ADM) formalism \cite{Arnowitt1959, Dewitt1968, Soo2014}. However, without considering the difficulties of a suitable quantization, this paper focuses on a possible classical-to-quantum \textit{bridge} between the SU$(1, 3)$ Yang--Mills gauge formalism and the gauge-like treatment of teleparallel gravity (Sect. \ref{sec:bridge}). \textcolor{black}{Drawing a parallel between SU(1,3) and SU(3) chromodynamics, we will say that the bridge defines a \textit{colored gravity}.}

\textcolor{black}{Henceforth, we can assume $M \cong \mathcal{M}$ to be the Minkowski spacetime, which is trivially a paralellizable non-compact manifold, and thus it admits a global cross-section (in the frame bundle) that derives a spinor structure. This is essential for constructing the SU(1,3)-colored connection, since this connection utilizes the spin structure to define parallel transport and covariant differentiation of spinor fields. At second order, the Minkowski metric is perturbed by small terms related to the spinor fields (Sect. \ref{sec:colored_metric}), but it is assumed to be negligible to define the spin structure.} 

\subsection{Classical-to-quantum bridge}
\label{sec:bridge}

To bridge the classical and quantum viewpoints of the coordinates, we have formulated a hypothesis on the spatial density of a particle located at the point $x\in M$ with 4-velocity $\mathbf{u} = \gamma_\mu u^\mu$. The volumetric density of a particle is a delta-type function,
\begin{equation}\label{eq:density0}
\wp(y)=\frac{1}{\sqrt{|g(y)|}}\delta^4(y-x),\qquad y\in M\,,
\end{equation}which contrasts to the quantum density operator $\hat \wp := \ket{\psi}\bra{\bar \psi}$ for Dirac spinors $\ket{\psi}$. For each subset $U\subset M$, the probability of finding the particle in $U$ is given by
\begin{equation*}
\varrho(y)=\int_U  \wp(y)\mathrm{d}\tau_x := \frac{1}{\sqrt{|g(y)|}}\int_U \delta^4(y - x)\mathrm{d}\tau_x = \begin{cases}
    1\;\;\text{if }y \in U \\
    0\;\;\text{if }y \notin U
\end{cases},
\end{equation*} where locally $d\tau_x=\sqrt{|g(x)|}d^4x \in\Omega^4(M)$. Therefore, the classical (point) density of probability current is simply $\wp \mathbf{u}$.

On the other hand, quantum field theory considers the current density from spinors (fields instead of point particles). Let $\psi$ be a Dirac spinor representing particles and $P$ be any operator. The expectation value is given by the $L^2$ inner product over $P$, \begin{equation} \label{eq:expectation}
    \braket{ P }_\psi = \mathfrak{Re} \braket{ \bar \psi  |\, P \, | \psi},
\end{equation} 
where $\bar \psi := \psi^{\dagger}\gamma_0$ is the Dirac adjoint in this Hilbert space. Classical quantum mechanics defines the density current as $u^\mu := \mathfrak{Re} \bra{\bar\psi} \hat{U}^\mu \ket{\psi}$ with the momentum operator $\hat{U}^\mu = \frac{\mathrm{i}}{m} \partial^{\mu}$ evaluated for some mass $m \in \mathbb{R}_{>0}$. Now, the components of the spacetime current $\mathbf{j} = \gamma_\mu j^\mu$ are $j^\mu := \bar \psi \gamma^\mu \psi$. Therefore, our hypothesis is that $\mathbf{j}$ should be classically equivalent to $\varrho\mathbf{u}$:
\begin{equation} \label{eq:bridge}
    \mathbf{j}  = \gamma_\mu \bar \psi \gamma^\mu \psi  \;\; \simeq  \;\;  \wp\,\gamma_\mu u^\mu   \;\; \Longrightarrow\;\;  \braket{ \gamma^\mu }_\psi =   \varrho|_{x}\, u^\mu = u^\mu.
\end{equation}where $x$ is the measured position of the ``classical particle'' and therefore $\braket{\hat \wp }_\psi = \braket{\mathbf{1}_4}_\psi  = \varrho|_{x} = 1$. Moreover, notice that Eq. \ref{eq:bridge} leads to \begin{equation} \label{eq:bridge2}
    \braket{ \mathbf{m}}_\psi = m \braket{ \mathbf{u}}_\psi = mu_\mu \braket{ \gamma^\mu }_\psi = m u_\mu u^\mu = m = \braket{ \rho }_\psi,
\end{equation} 
where $\rho := \varrho m$ is the classical point energy density and $\mathbf{u} = \gamma^\mu u_\mu = \gamma_\mu u^\mu$. From now onwards, we use the symbol ``$\simeq$'' to refer to the classical-quantum bridge:\begin{equation} \label{eq:bridge3}
  a \simeq b \;\;\;\; \Longleftrightarrow  \;\;\;\;  \braket{a}_\psi = \braket{b}_\psi\;, 
\end{equation}which has the property that $C\,a \simeq C\,b$ for any $C \in \mathbb{R}$. Therefore, one recovers the (quantum) momentum operator $\mathrm{i}\partial_\mu \simeq m_\mu$ and the Dirac equation for $\psi = \psi_0 \exp(-\mathrm{i}\, m \gamma_\nu dx^\nu)$, \begin{eqnarray}\label{eq:dirac0}
    \mathrm{i}\gamma^\mu \partial_\mu \simeq m\,,
\end{eqnarray}or, similarly, $\mathrm{i}\gamma^\mu \partial_\mu - m \simeq 0$. 

\subsection{Perturbed spinor frames}
\label{sec:perturbedspinor}

Let $\Psi= \{\psi_n\}_{n=1}^4 =\{\psi_1,\psi_2,\psi_3,\psi_4\}$ be a set of 4 Dirac spinors on a $M$, that is, a set of 4 column vectors of fermions or pairs of Weyl spinors. Recall that each Dirac spinor has 4 components $\psi_n(\psi_n^\mu)_{\mu=0}^3$; therefore, they can be represented by matrices $\ket{\Psi} \in \pazocal{M}_4(\mathbb{C}) = \mathbb{C}\times Cl_{1,3}(\mathbb{R},g)$, and the total spin of $\Psi$ is an integer value between -2 and 2. These spinor fields represent \textit{particles} with the same energy, $m \in \mathbb{R}_{>0}$, ideally expressed by the 4-momentum $\mathbf{m}$, i.e., a matrix satisfying $\mathbf{m}\bigcdot\mathbf{m} = m^2\mathbf{1}_4$, where $\mathbf{1}_4$ is the identity matrix. The moment can be decomposed into components using the generators of the spacetime algebra, $\{ m_\mu := \mathbf{m} \bigcdot \gamma_\mu \}_\mu$, thus yielding the set of angular wavenumbers of $\Psi$. As a physical assumption, we identify these angular wavenumbers with classical momenta, 
\begin{equation}\label{eq:classicalangularmomenta}
\mathbf{m}\bigcdot\gamma_\mu=m_\mu= m u_\mu \mathbf{1}_4 := m \frac{dx_\mu}{d\tau}\mathbf{1}_4.
\end{equation}
Therefore, the velocity is $\mathbf{u} := \mathbf{m}/m = \gamma^\mu u_\mu$. In addition, the spacetime element is given by $\mathbf{d}\boldsymbol{\tau}:=\gamma_{\mu} dx^\mu$, whence $\mathbf{m}\bigcdot \mathbf{d}\boldsymbol{\tau} =  m_{\mu} dx^\mu$. As a result, we may write \begin{equation} 
\Psi := \Psi_0(\mathbf m) \exp \left(- \mathrm{i} \int \mathbf{m}  \bigcdot \mathbf{d}\boldsymbol{\tau}\right). \end{equation}From now on, the integration symbol $(\int)$ is omitted 
\begin{equation}
\Psi\simeq\Psi_0(\mathbf{m}) \exp \left(- \mathrm{i} \, \mathbf{m}  \bigcdot \gamma_\mu\, dx^\mu \right).
\end{equation} Moreover, we use assume a time-dependent Ansatz for the spinor field to separate the phase-coherent spinor field $\Psi_0(\mathbf{m})$ from the time-dependent scalar field $\xi(\tau) = \exp \left(- \mathrm{i} \,\mathbf{m}  \bigcdot \mathbf{d}\boldsymbol{\tau}\right)$.

 Let $\mathbf{A} = {A}_\mu\gamma^\mu$ be a gauge potential field, expressed using the spacetime algebra \cite{Dressel2015}, while let $A=A_\mu dx^\mu\in\Omega^1(M,\mathfrak{su}(1, 3))$ be a connection on the principal SU(1, 3)-bundle defined in Sec. \ref{sec:complex} over $M$, whose components $A_\mu = A_\mu^I\lambda_I$ are represented by using a basis $\{\lambda_I\}_{I=1}^{15}$ of $\mathfrak{su}(1, 3)$.

Moreover, let ${\partial}_\mu  \;\; \mapsto \;\; {\nabla}_\mu := {\partial}_\mu - \mathrm{i}q{A}_\mu$ be the transformation of the linked covariant derivative. In coordinates, this is equivalent to the introduction of a phase $\varphi(x)$ into the spinor field given by the appearance of the potential field $\mathbf{A}$. According to Eq.~\ref{transfA} with vanishing starting potential, i.e. $\nabla_\mu^{(init)} = \partial_\mu$, the final potential field is ${A}_\mu(x) = q^{-1}[\nabla_\mu^{(init)},  \varphi(x)] = q^{-1}[ \partial_\mu,  \varphi(x)] = q^{-1} {\partial}_\mu \varphi(x)$. Therefore, the phase is $\varphi(x) = \lambda_I\varphi^I(x) = q  A_\mu^I(x)\lambda_I d x^\mu = q A_\mu(x) d x^\mu$, and is added to transform the field by the unitary operator $U(x) := \exp(\mathrm{i}\varphi(x)) = \exp({\mathrm{i}q A_\mu(x) d x^\mu})$. To compensate this phase, the covariant derivative acts like the Hermitian adjoint operator  $U^{\dagger}(x) := \exp({-\mathrm{i}\varphi(x)}) = \exp({-\mathrm{i}q A_\mu(x) d x^\mu})$ as follows  
\begin{equation}
\Psi  \; \; \mapsto  \; \;  \hat\Psi: = U^{\dagger}(x) \Psi = \Psi_0(\mathbf{m})\exp \left(-\mathrm{i}\left( \mathbf{m} \bigcdot \gamma_\mu + q A_\mu \right) dx^\mu\right).
\end{equation} 
It can be easily verified that the gauge covariant derivative ${\nabla}_\mu = {\partial}_\mu - \mathrm{i}q{A}_\mu$ satisfies the symmetric transformation $\hat\Psi^\dagger {\hat\nabla}_{\mu }  \hat\Psi = \Psi^\dagger {\nabla}_{\mu }  \Psi$, with ${\hat\nabla}_{\mu } =  U({x}) {\nabla}_{\mu }U^{\dagger }({x})$. The operator $U^{\dagger}(x) = \exp({-\mathrm{i}q A_\mu(x) dx^\mu}) \approx \exp({-\mathrm{i}q\int A_\mu(x) d x^\mu})$ is also related to the Wilson loop, a generalization of the Aharanov-Bohm effect \cite{Ferrari1989, Alfonsi2020}. For a single-valued wavefunction, the phase change around the loop must be an integer multiple such as $\varphi(x) = 2\pi\,n$ with $n \in \mathbb{N}$, but these quantization aspects are not addressed in this paper.

A key point is that the compensatory phase $\varphi({x}) = qA_\mu(x) dx^\mu = q(\mathbf{A}(x)  \bigcdot \gamma_\mu) dx^\mu = q \mathbf{A}(x)  \bigcdot \mathbf{d}\boldsymbol{\tau}$ yields the following transformation,
\begin{eqnarray} \label{eq:gammaandmass}
(\gamma_\mu \bigcdot  \mathbf{m}) dx^\mu     &  \mapsto  & 
\left(\gamma_\mu + q{A_\mu}\bigcdot\frac{\mathbf{m} \,}{m^2} \right) \bigcdot \mathbf{m}\, dx^\mu,
\end{eqnarray}
where we have used that $\mathbf m\bigcdot\mathbf m=m^2\mathbf 1_4$. Moreover, we denote by $\hat \gamma_\mu$ the perturbed spacetime generators,
\begin{equation} \label{eq:gamma_matrix}
{\gamma_\mu}  \; \;   \mapsto   \; \; {\hat \gamma_\mu} := {\gamma_\mu}  +  {A_\mu}\bigcdot\frac{q}{\mathbf{m}} =  \gamma_\mu  + a_\mu^I \lambda_I\bigcdot \frac{\mathbf{m}}{m},
\end{equation}
where we denote $a_\mu^I := qA_\mu^I/m$ and $1/\mathbf{m} := \mathbf{m}/m^2$ for simplicity. Here, the four-vector $A_\mu^I$ has one timelike and three spacelike components, analogous to a four-velocity. On the other hand, it is possible to interpret the compensatory phase of Eq. \ref{eq:gammaandmass} as an energy perturbation, $\mathbf{m} \; \mapsto \;  \mathbf{\hat m} := \mathbf{m} + q\mathbf{A}$. This is equivalent to Eq. \ref{eq:gamma_matrix}, since $\gamma^\mu\bigcdot \mathbf{m} \bigcdot \hat \gamma_\mu = \mathbf{m} +q\mathbf{A} = \mathbf{\hat m}$. Now, we can define $\mathbf{\hat u} := \mathbf{\hat m}/m = \gamma^\mu \hat u_\mu = \mathbf{u} + q\mathbf{A}/m$, where \begin{equation} \label{eq:velocity}
u_\mu \;\; \mapsto \;\; \hat u_\mu :=  \mathbf{u} \bigcdot \hat \gamma_\mu = \mathbf{1}_4 u_\mu + A_\mu \frac{q}{m} \;\; \in\mathbb R\oplus\mathfrak{su}(1, 3),
\end{equation} with $A_\mu = A^I_\mu \lambda_I$. Here, $\mathbb R$ is regarded as a subspace of $\mathfrak{gl}(4,\mathbb C)$ via the injection $a\mapsto a\mathbf 1_4$. Note that Eqs. \ref{eq:gamma_matrix} and \ref{eq:velocity} satisfy the bridge defined by Eqs. \ref{eq:bridge} and \ref{eq:bridge2}. The velocity component $\hat u_\mu$ is an element of $\mathbb R\oplus\mathfrak{su}(1, 3)$ because $\lambda_I\in \mathfrak{su}(1, 3)$  and $A^I_\mu\, ,q/m\, ,u_\mu \in \mathbb{R}$. Therefore, the velocity $\hat {\mathbf{u}} = \gamma^\mu \hat u_{\mu}$ implies that the translation $\delta \hat {\mathbf{x}} = \hat{\mathbf{u}}\, \delta \tau$ is an element of $\mathbb R^{1,3} \oplus \mathfrak{u}(1, 3)$, which includes the Poincaré algebra $\mathbb R^{1,3} \oplus \mathfrak{o}(1, 3)$. Thus, all the above calculations performed with $A_\mu$ can be replicated for any element of $\mathbb R^{1,3} \oplus \mathfrak{u}(1, 3)$.


Moreover, notice that the perturbation of the 4-velocity is a \textit{current} term $q A_\mu/{m}$, similar to the Ferrari proposal \cite{Ferrari1989} and the \textit{London equation} in electrodynamics \cite{Frohlich2013, BadiaMajos2006}, interpreted as the ground state of the tetrad under the selected gauge, consistent with the Dirac equation. Up to first order, by applying the change of $0 \mapsto A_{\mu}$ to the co-tetrad transformation $e^a \mapsto \hat e^a$ (Eq.~\ref{eq:tetrad}), and by assuming local invariance of the spacetime algebra, ${\gamma^a} :=  e_{\mu}^{\ a} {\gamma^\mu} \equiv \hat e_{\mu}^{\ a}\bigcdot{\hat \gamma^\mu}$, Eq. \ref{eq:gamma_matrix} yields
\begin{equation*}
e_{\mu}^{\ a} {\gamma^\mu} = \hat e_{\mu}^{\ a}\bigcdot{\hat \gamma^\mu} \approx \left( e_{\mu}^{\ a} + \phi_{\mu}^{\ a} \right)\bigcdot\left({\gamma^\mu} - {A^\mu} \frac{q}{\mathbf{m}} \right) 
\Longrightarrow \quad\phi_{\mu}^{\ a}\bigcdot{\gamma^\mu} \approx e_{\mu}^{\ a}{A^\mu}\bigcdot\frac{q}{\mathbf{m}} = {A^a}\bigcdot\frac{q}{\mathbf{m}},
\end{equation*}where we write $A^a := e_{\mu}^{\ a}{A^\mu}$. The co-tetrad perturbation $\phi_{\mu}^{\ a}$ \textbf{is now a matrix} that is not unique. We use the following as an ansatz, 
\begin{equation}\label{eq:phi_perturbed}
 \phi_{\mu}^{\ a} := -\mathrm{i}q {A}_\mu^{\ a} := q A^a  \frac{m_\mu}{m^2} =  \frac{q}{m} A^a  u_\mu = A^a\bigcdot\frac{q}{\mathbf{m}}\bigcdot\gamma_\mu,
\end{equation}
where $A_\mu^{\ a}:=i A^am_\mu/m^2$ and $m_\mu = m u_\mu \mathbf 1_4$, being $u_\mu := dx_\mu/d\tau$ the 4-velocity covector (recall Eq. \ref{eq:classicalangularmomenta}). The constant $-\mathrm{i}q$ is explicitly stated for convenience. Since $e_a(e^b)\mathbf 1_4 = \hat e_a(\hat e^b) = \hat e_{\ a}^\mu\bigcdot\hat e^{\ b}_\mu = \delta^a_b\mathbf 1_4$, the tetrad perturbation is $\phi_{\ a}^\mu = e^{\mu}_{\ b}e^{\nu}_{\ a}\phi_{\nu}^{\ b} \approx  -\mathrm{i}q {A}_{\ \,a}^\mu$, with $\hat e^\mu_{\ b} = e^\mu_{\ a} - \phi_{\ a}^\mu$. To this, $ {A}_\mu^{\ a} = e_{\mu}^{\ b}e_{\nu}^{\ a} {A}^\nu_{\ b}$ is implicitly related to the teleparallel gauge-like potential $\phi_\mu = \phi_\mu^{\ a}\partial_{a}$, which implies that 
\begin{equation}\label{eq:fundamental}
A_a \simeq  {A}^\mu_{\ a} \partial_\mu = A_a\bigcdot \frac{\mathrm{i}}{\mathbf{m}}\bigcdot\gamma^\mu  \partial_\mu \,.
\end{equation} 

The expression of the gauge potential (Eq.~\ref{eq:fundamental}) is fundamental for the proposed translation gauge $\phi_{\mu}^{\ a} = - \mathrm{i}q {A}_\mu^{\ a}$. However, the Dirac-like equation (see Eq. \ref{eq:dirac0}), $\frac{\mathrm{i}}{\mathbf{m}} \bigcdot  \gamma^\mu \partial_\mu \simeq \mathbf{1}_4$, is only valid for expectation values of the spinor $\Psi = \Psi_0(\mathbf m)\exp({-\mathrm{i}\, \mathbf{m}\bigcdot \mathbf{d}\boldsymbol{\tau}})$ selected, i.e.,
\begin{equation*}
 \frac{\mathrm{i}}{m} u^\mu \braket{\partial_\mu}_\Psi = 1.
\end{equation*}
For other cases, it is necessary to consider the replacement $\mathbf{m} \leftrightarrow \mathrm{i}\gamma^\mu \partial_\mu\tilde\Psi$, as it corresponds to a new field $\tilde\Psi$. 

\subsection{Colored spacetime algebra for spinors}

Given the set of spinors $\Psi(x) = \Psi_0(\mathbf m) \exp({-\mathrm{i}\, \mathbf{m}\bigcdot \mathbf{d}\boldsymbol{\tau}})$, we define an \textit{extended teleparallel gauge} (ETG) translation of the coordinates, 
\begin{equation}
    \delta x^a \approx - \left(\, \mathrm{I}_4 \,\phi^{\ a}_{\nu}{}_{scalar}  + \phi^{\ a}_{\nu} \,\right)\delta x^\nu
\end{equation}
where $\phi^{\ a}_{\nu}{}_{scalar}$ is the usual TEGR perturbation, $\phi^{\ a}_{\nu}$ is the perturbation developed in the section above and, therefore, $\delta x \in  \mathbb R^{1,3} \oplus \mathfrak{u}(1, 3)$. To explore possible limits of electromagnetism, we neglect the scalar term $\phi^{\ a}_{\nu}{}_{scalar} \approx  0$. 

Now, $\delta x^a \approx  - \phi^{\ a}_{\nu}\delta x^\nu \simeq \mathrm{i}q {A}^{\ a}_{\nu} \delta x^\nu$ is related to the U$(1,3)$ gauge transformation and the Wilson line, modifying the trajectories and therefore also the (oscillating) local coordinates $\{x^\mu\}$, which become \textbf{matrices}. This ETG relation yields the following transformations:
\begin{eqnarray} \label{eq:gamma_hat}
 {\gamma_{\mu}} \hspace{3mm} & \mapsto & \hspace{5mm}   \hat{\gamma}_{\mu} :=  {\gamma_{\mu}} + A_\mu\bigcdot\frac{q}{\mathbf{m}}, \\ \nonumber 
d{x}^\mu \hspace{3mm} & \mapsto & \hspace{5mm} \hat{d{x}^\mu} \approx  \mathbf{1}_{4} dx^\mu - A^\mu\bigcdot\frac{q}{\mathbf{m}} \bigcdot \gamma_\nu dx^\nu \; \simeq \;  \mathbf{1}_{4} dx^\mu + \mathrm{i}q {A}^{\ \mu}_{\nu}  dx^\nu,  \\
 \label{eq:transf_partial}
\partial_\mu  \hspace{3mm} & \mapsto & \hspace{5mm}  \hat\partial_\mu \approx   \mathbf{1}_{4} \partial_\mu + A_\mu\bigcdot\frac{q}{\mathbf{m}} \bigcdot \gamma^\nu \partial_\nu \; \simeq \;   \mathbf{1}_{4} \partial_\mu - \mathrm{i}qA_\mu.
\end{eqnarray}
where 
\begin{equation*}
A_\mu  = {A}_\mu^{\ a} \partial_a \simeq  A_\mu\bigcdot\frac{\mathrm{i}}{\mathbf{m}} \bigcdot \gamma^a \partial_a,\qquad\frac{q}{\mathbf{m}} \bigcdot \gamma_\nu = \frac{q}{m^2}\, m_\nu = \frac{q}{m} \, u_\nu\mathbf 1_4.   
\end{equation*}
In particular, the new coordinates satisfy
\begin{equation*}
{\hat\gamma_{\mu}}\bigcdot{\hat\gamma_{\nu}}\bigcdot\hat{d{x}^\mu}\bigcdot\hat{d{x}^\nu}=\boldsymbol{d\tau}\bigcdot \boldsymbol{d\tau}={\gamma_{\mu}} \bigcdot {\gamma_{\nu}}d{x}^\mu d{x}^\nu.
\end{equation*}
Moreover, the following must be satisfied: $\mathbf{1}_4 = \mathbf{1}_4 \partial_{\mu}dx^{\mu} = \hat\partial_{\mu}\bigcdot\hat{dx^{\mu}}$, from which we deduce that $(A_\mu u^\nu)\partial_\nu  dx^{\mu} = \partial_\mu(A^\mu u_\nu \,  dx^\nu)= (A^\mu \partial_\mu dx^\nu)  u_\nu$ with $u_\nu \partial_\mu dx^\nu = u_\nu \partial_\mu u^\nu d\tau = 0$ and $\partial_\mu A^\mu = 0$.  Applying the derivative $\hat{\partial}_{\mu}$ to our particular set of spinors, $\Psi(x)$, we obtain $\hat{\partial}_{\mu}\hat\Psi \approx  - \mathrm{i}(m_\mu + q A_\mu ) \Psi \;\; = \nabla_\mu \Psi $. Naturally, condition $\hat{\partial}_{\mu}\hat\Psi = {\partial}_{\mu} \Psi$ is satisfied, which is equivalent to $\hat\partial_{\mu}\bigcdot\hat{dx^{\mu}} = \partial_{\mu}dx^{\mu}$, where the new coordinates behave like matrices. In other words, the transformed spacetime basis for a particle-field $\Psi$, $(\hat{dx^{\mu}},~\hat\partial_{\mu})$, is equivalent to the introduction of a (non-abelian) Wilson phase, $-\varphi = -A_\mu dx^\mu$, to the field while keeping the original basis ($dx^{\mu}$, $\partial_{\mu}$). 

Observe that the \textit{London equation current} with gauge field $A^\mu$ in U$(1, 3)$ transforms the spacetime coordinates as follows:
\begin{equation}
    \hat x^\mu \;  \approx\;   \mathbf{1}_{4} x^\mu - \int \frac{q}{m} A^\mu u_\nu dx^\nu =    \mathbf{1}_{4} x^\mu - \int \frac{q}{m} A^\mu d\tau \;\; \in\mathbb{R}\oplus\mathfrak{u}(1, 3)
\end{equation} 
Hence, $\hat x^\mu \;  \approx\;\left(\eta^{\mu\nu} - q A^\mu u^\nu/m \right)x_\nu$. 

Moreover, the 4-velocity transforms as
\begin{eqnarray}
    \hat u^\mu \;  \approx\;      \mathbf{1}_{4} u^\mu -  \frac{q}{m} A^\mu. 
\end{eqnarray}
We conclude from Eq. \ref{eq:velocity} that it satisfies $\hat u^\mu\bigcdot\hat u_\mu =\mathbf 1_4$.

The matrix structure of the 4-velocity $\hat u^\mu$ and spacetime coordinates $\hat x^\mu$ can be interpreted, respectively, by the classical-to-quantum bridge to the $m$-normalized momentum operator $\hat U^\mu := \frac{1}{m} \hat P^\mu := \frac{\mathrm{i}}{m}\left(\mathbf{1}_{4} \partial^\mu +  q A^\mu\right)$ and the corresponding translation-position operator $\hat X^\mu$.

\subsection{Colored metric}
\label{sec:colored_metric}

The perturbed \textit{metric coordinates} evaluated at $x$ for $\Psi(x)$ are given by the perturbed spacetime generators (Eq. \ref{eq:gamma_hat}). Taking $\hat e_\mu := \hat e_{\mu}^{\ a}\partial_a = e_\mu + \phi_{\mu}$, the metric components are now matrices themselves:
\begin{equation}\label{firstordermetric}
\begin{array}{ccccc}
g_{\mu\nu} & := & \hat\gamma_\nu\bigcdot\hat\gamma_\nu & = & \hat e_\mu\bigcdot\hat e_\nu\\
& \approx & \eta_{\mu\nu}\mathbf 1_4+2\displaystyle\frac{q}{m} A_{(\mu} u_{\nu)} & \approx & \eta_{\mu\nu}\mathbf 1_4+2\phi_{\mu\nu},
\end{array}
\end{equation}
where $A_{(\mu} u_{\nu)} := (A_{\mu} u_{\nu} +  A_{\nu} u_{\mu})/2$ and $\phi_{\mu\nu}=\phi_{(\mu}\bigcdot e_{\nu)}$. The inverse is given by
\begin{equation*}
g^{\mu\nu}=(g^{-1})^{\mu\nu}\approx\eta^{\mu\nu}\mathbf 1_4-2\frac{q}{m}A^{(\mu}u^{\nu)}\approx\eta^{\mu\nu}\mathbf 1_4-2\phi^{\mu\nu}.
\end{equation*}
Alternatively, the metric components can be written as $g_{\mu\nu}\approx\eta_{\mu\nu}\mathbf 1_4-\hat u_\mu\bigcdot\hat u_\nu+u_\mu u_\nu\mathbf 1_4$ and $g^{\mu\nu}\simeq\eta^{\mu\nu}\mathbf 1_4+\hat u^\mu\bigcdot\hat u^\nu-u^\mu u^\nu\mathbf 1_4$. Observe that $\hat e_\mu$ and $e_\mu$ behave as tetrads of directions or \textit{four-velocities} in the corresponding algebra and frame, while $\phi_{\mu\nu}$ is identified as a linearized perturbation. In particular, the coordinates and the metric components are complex matrices, as the perturbation generators $\{A_\mu u_\nu\}_{\mu,\nu}$ are not elements of $Cl_{3,1}(\mathbb{R})$. Moreover, now, the metric represents a \textit{spacetime-particle field} instead of the classical spacetime metric.

\vspace{3mm}

\begin{definition} \label{def:simplygravity} \textbf{(Simply gravity)}. Let $g_{\mu\nu} := \mathbf{1}_4 \eta_{\mu\nu}  + \sum_r \varsigma_r \, \hat u_{\mu}^r \hat u_{\nu}^r$ be the metric components with perturbation such that $ (\hat u_{\mu}^r \hat u_{\nu}^r)  : M \,\to \,(\mathbf{1}_{4}\mathbb{R} \oplus  \mathfrak{u}(1, 3)) \otimes  (\mathbf{1}_{4}\mathbb{R} \oplus  \mathfrak{u}(1, 3))$ is multiplied by a scalar field $\varsigma_r : M \to \mathbb{R}$. The effects of the perturbation will be named ``simply gravity'' whenever each perturbation component is a purely diagonal matrix $(\hat u_{\mu}^r \hat u_{\nu}^r)(x) \in \mathbf{1}_{4}\mathbb{R}$, that is, $\varsigma_r \,\hat u_{\mu}^r \hat u_{\nu}^r = \mathbf{1}_4 \varsigma_r \,k_{\mu}k_{\nu}$ for some Kerr--Schild unitary vector field $k = (k_{\mu})_\mu : M \to \mathrm{T}M$. Moreover, if $\eta_{0,0}\,\varsigma_r \le 0$ the metric represents a usual (simply) \textbf{gravity spacetime}, while for $\eta_{0,0}\,\varsigma_r > 0$, it leads to a (simply) \textbf{gravity source}, corresponding to ``matter sources'' instead of ``effects on spacetime curvature''.
\end{definition}

\vspace{2mm}

\textcolor{black}{\begin{remark}[\textbf{Spacetime}] We assume that spacetime is perturbed at second order, $M \to 
\hat M = M \oplus O(\hat u^2)$, as a result of the small terms added to the metric $\mathbf{1}_4 \eta_{\mu\nu} \to g_{\mu\nu}$. \end{remark}}

\vspace{3mm}

\begin{remark}[\textbf{Geons}] The idea of ``geons'' conceptualized by \cite{Wheeler1961} is very similar to the simply gravity sources defined above.
\end{remark}

\subsection{Double-copy gauge potential}
\label{sec:doublecopy}

To derive more interesting effects, it is necessary to analyze the second-order terms. In particular, one can pay attention to the definition of the origin of the SU(1, 3) gauge potential following the covariant Lienard-Wiechert or Cornell-like forms. The corresponding perturbation of the spacetime algebra is given by\begin{equation}
\label{eq:gamma_matrix2}
\gamma_\mu\; \;   \mapsto   \; \; \hat\gamma_\mu := \gamma_\mu +  \kappa\, {A_\mu}\bigcdot\mathbf{A} (0) \, , 
\end{equation} where the origin (conveniently normalized with the quadratic Casimir operator),
\begin{equation*}
\mathbf{A} (0) =  \frac{q}{\kappa \mathbf{m}} = \frac{q}{\kappa m^2}\,\mathbf{m} \; \in \mathcal{M}_4(\mathbb{C}),
\end{equation*}
is a matrix proportional to the momentum $\mathbf{m} = m^\mu \gamma_\mu = m_\mu \gamma^\mu $. If we denote by $\mathbf{\hat A}  := \mathbf{A} - \mathbf{A}(0)$ the relative gauge with coordinates $\hat A_{\mu} := 
\mathbf{\hat A} \bigcdot \gamma_\mu$, then the second terms of the metric are found as above, but using the double gauge copy of $\hat A$ instead (recall Eq. \ref{firstordermetric}): \begin{equation} \label{eq:exactsolution}
\begin{array}{ccccl}
g_{\mu\nu} & = & \hat \gamma_\mu\bigcdot\hat \gamma_\nu & = & \eta_{\mu\nu} - \kappa \hat A_{\mu}\bigcdot\hat A_{\nu} +  \kappa A_{\mu}(0)\bigcdot A_{\nu}(0) \\
& &  & = & \eta_{\mu\nu}  - \kappa A_{\mu}\bigcdot A_{\nu} +  2q A_{(\mu}   u_{\nu)}/m,
\end{array}
\end{equation} where $A_\mu = \mathbf{A} \bigcdot \gamma_\mu$ is now a SU(1, 3) gauge potential (boson) and the last term represents a \textbf{gravity source} at $x = 0 \in \mathbb{R}^4$ (see Def. \ref{def:simplygravity}), while the first perturbation term, $A_{\mu}\bigcdot A_{\nu}$, corresponds to a \textbf{gravity spacetime} linked to a pair of bosons entangled  (i.e., a candidate for \textit{graviton}). 


Since $A_\mu\in\mathfrak{su}(1, 3) \subset \mathfrak{u}(1, 3)$ and $u \in \mathbb{R}^{1,3}$, perturbations of the complexified metric 
\begin{equation*}
g\sim\eta+\hat A\otimes\hat A=\eta+(u\oplus A)\otimes(u\oplus A)
\end{equation*}
can be identified, via the canonical isomorphisms of tensor products, with elements of the $\left( \mathbb{R}^{1,3}\oplus\mathfrak{u}(1, 3) \right) \otimes \left(\mathbb{R}^{1,3}\oplus\mathfrak{u}(1, 3) \right)$ space, instead of the classical $ \mathbb{R}^{1,3} \otimes  \mathbb{R}^{1,3}$ tensor space.


Observe that when the perturbation components are purely diagonal, the metric provides a $Cl_{4}(\mathbb{C})$ algebra generated by $\{\hat \gamma_\mu\}_{\mu = 0}^3$. As a particular case, one can recover the classical double-copy used in the Kaluza--Klein and the Kerr--Schild--Kundt metrics, $g\sim\eta+A\otimes A$ (e.g., the black holes of Kerr--Newman and Reissner--Nordstr\"om) \cite{Monteiro2021}, and can consider higher-order corrections to the Lagrangian (see Sect. \ref{sec:lagrangian}) by following analogies with the Born--Infeld / D-Brane actions \cite{Kogan2003}. Similarly, the dust-based linearized metric, $g \sim \eta + u \otimes u$, can be recovered as a limiting case of the torsion-based colored gravity metric. 

Schematically, the torsion of the developed metric describes a double-helix-like structure consisting of pairs $A \otimes A$ of entangled $\mathfrak{su}(1,3)$ vector fields or \textit{bosons} (e.g. QED-carrier photons or QCD-carrier gluons). Geometrically, these potential fields are interpreted as the connection of a double-copy gauge transformation generated by an extended Poincaré algebra $\mathbb{R}^{1,3} \oplus \mathfrak{u}(1,3)$. From a physical perspective, these bosons are virtual particles of exchange in each interaction (transformation) of the extended Poincaré symmetry group.

The features of the double-gauge metric are very rich and deserve to be analyzed in detail in some future work. The simplest case is the abelian U(1) electrodynamics, which is developed in \textbf{Sect. \ref{sec:geometric}}.


From the compatibility hypothesis on both sides of the metric perturbation $(A, u)$, we conclude that the natural algebra is built with $\mathfrak{su}(1, 3)$. The idea of a double-copy gauge potential $A\otimes A \in \mathfrak{su}(1, 3) \otimes \mathfrak{su}(1, 3)$ is linked to some remarkable isomorphisms. For instance, the chiral group for spinors is $\mathrm{SU}(2) \times \mathrm{SU}(2)$. The representatives of the spacetime algebra are isomorphic to $2\times 2$ quaternion matrices, ${M}_{2}(\mathbb{H})$, and to the double $2 \times 2$ unitary matrices, $U(2)$, that is, $Cl_{1,3}(\mathbb{R}) \simeq \mathfrak{gl}(2, \mathbb{H}) \simeq \mathfrak{u}(2) \oplus \mathfrak{u}(2)$. On the other hand, the quaternion formulation of the GR provides interesting links to electrodynamics \cite{Sachs1968, Crater2011}.

The complexification of the spacetime algebra leads to the Dirac algebra, $Cl_{4}(\mathbb{C}) \simeq Cl_{1,3}(\mathbb{R}) \otimes \mathbb{C} \simeq \mathfrak{u}(2) \oplus \mathfrak{u}(2) \oplus \mathfrak{u}(1)$, which contains the electroweak interaction algebra, $\mathfrak{su}(2) \oplus \mathfrak{u}(1)$. Moreover, the reoriented elements of $Cl_{1,3}(\mathbb{R})$ representing the generators of SU(3) bring the well-known degree-preserving transformations of the Lorentz group and comprise enough structure for both the color- and the flavor-SU(3) \cite{Schmeikal2001, Schmeikal2004}. Nevertheless, the Dirac-algebra generators can be used to linearly span the whole algebra $\mathfrak{gl}(4,\mathbb{C})$, that is, any $4 \times 4$ complex matrix can be expressed by using the sixteen Dirac matrices as a basis \cite{RedKov2008, VazRocha2016}. The $\mathfrak{su}(4)$ algebra allows to model quarks and leptons within the same symmetry \cite{MarschandNarita2015, BarbieriandTesi2018}, while $\mathfrak{u}(1,3)$ is important to describe strong interactions of quarks and gluons in a non-perturbative domain at large interaction distances \cite{Khruschev2004}. In this context, it is conceivable that the $\mathfrak{su}(1,3)$ algebra also leads to a similar quark-lepton model, as happens in the $\mathfrak{su}(4)$ case, and since it has similarities with Wess-Zumino-Witten models in 2 dimensions and with Chern-Simons theories in 3 dimensions \cite{Margolin1992, Tseytfin1995}.




Among the relevant subalgebras of $\mathfrak{su}(1, 3)$, we enumerate $\mathfrak{su}(3)$ and $\mathfrak{su}(2)\oplus \mathfrak{u}(1)$, which are reductive of maximal rank. However, $\mathfrak{su}(3) \oplus \mathfrak{su}(2) \oplus \mathfrak{u}(1)$ cannot be embedded into $\mathfrak{su}(1, 3)$, although it admits an embedding into the double $\mathfrak{su}(1, 3)$ and double $\mathfrak{su}(4)$, or more generally double $\mathfrak{u}(4)$ \cite{RedKov2008, Castro2012, CembranosDiezValle2019}. Therefore, the (1, 3)-\textit{color} symmetries can be broken into the \textit{3-color} and the \textit{hypercharge} symmetries of the Standard Model, thanks to the coupling factors and the partial composition of the Higgs boson \cite{Schmeikal2004, CembranosDiezValle2019, Almeida2003, Ferreti2014, Cossu2019, Gertov2019}. The role of this breaking is played by the perturbation properties of $Cl_{4}(\mathbb{C})$. The solution to obtain compatibility of the above mentioned facts is to describe the perturbation of the spacetime algebra using either the Lie algebra $\mathfrak{su}(1, 3) \oplus \mathfrak{su}(1, 3)$ or $\mathfrak{u}(1, 3) \oplus \mathfrak{u}(1, 3)$.
   
\subsection{Colored covariant derivative}

The covariant derivative for a scalar field is $D_\mu := \hat e_\mu^{\ a}\partial_a$, which is given by the co-tetrad components $\hat e_\mu^{\ a}$ as representatives of the teleparallel translational gauge-like field $\phi_{\mu}^{\ a}$. The (non-trivial) co-tetrads and tetrads are given by
\begin{eqnarray} \label{eq:cotetrad1} 
\hat e^a = \hat e_{\mu}^{\ a} dx^{\mu} \approx  e^{a} +  \phi_{\mu}^{\ a} dx^{\mu} \approx e^a + \phi^a, \;\;
\\
\label{eq:tetrad1}
\hat e_a =  \hat e^{\mu}_{\ a} \partial_{\mu} \approx  e_a - \phi^{\mu}_{\ a} \partial_{\mu} \approx  e_a - \phi_{a}, \;\; 
\end{eqnarray} respectively, where the identities $\phi_{a} = \phi^{\mu}_{\ a} \partial_{\mu}$ and $\phi^{a} = \phi_{\mu}^{\ a}dx^{\mu}$ have been used, and the identity matrices have been omitted for brevity. For a spinor field $\Psi$ expressed in the above time-dependent Ansatz, total covariant derivative is $D_\mu = \hat e_\mu^{\ a}\partial_a + \omega_\mu$, where $\omega_\mu$ is the spin connection. Considering weak gravitational field, the spin connection components are negligible (because they are derivates of $\phi$), minimizing the effects of local Lorentz transformations on the spinor field. Therefore, we assume that the primary contributions to the covariant derivative come from the translational gauge field $\phi$.

In terms of the potential field $A_\mu$ of the SU(1,3) gauge, the Eqs. \ref{eq:cotetrad1} and \ref{eq:tetrad1} read 
\begin{eqnarray} \label{eq:cotetrad3} 
\hat e^{a} \approx e^a + \phi^a \approx e^a + \frac{q}{m}  A^a  u_\mu dx^{\mu}  \simeq e^a - \mathrm{i}q {A}^{\ a}_{\mu} dx^\mu, 
\\ \label{eq:tetrad3}
\hat e_a \approx  e_a - \phi_{a}  \approx  e_a - \frac{q}{m} A_a  u^{\mu} {\partial}_\mu \simeq e_a + \mathrm{i}q A_a.
\end{eqnarray} 
It follows that the Cartan covariant derivative $D_\mu$ coincides with the teleparallel gauge
\begin{eqnarray} \label{eq:covariant3}
D_\mu \equiv\hat\partial_\mu \equiv \hat e_\mu = \hat e_\mu^{\ a}\partial_a \approx  e_\mu + \frac{q}{m} A_\mu  u^a \partial_a  \simeq  \partial_\mu - \mathrm{i}q A_\mu = \nabla_\mu.
\end{eqnarray} 
Finally, the \textit{expectation} equivalence $\nabla_\mu \simeq D_\mu$ between the SU(1, 3) gauge derivative and the teleparallel derivative can be set for each \textit{particle-field} involved (expectation value). This equivalence is of major relevance to ensure the consistence of the theory for non-abelian fields $\mathbf{A}$.

\subsection{Colored Gravity Lagrangian density}\label{sec:lagrangian}

Due to the equivalence of the covariant derivatives, the 2-form torsion tensor yields the field strength of the teleparallel translational gauge (Eq. \ref{eq:strength_transl}), which behaves like a 2-form curvature. Therefore, by introducing the perturbed co-tetrads (Eq. \ref{eq:cotetrad3}) in the field strength and by applying expectation values (Eq. \ref{eq:bridge3}), we obtain the SU(1, 3) field strength (geometrically, the curvature 2-form): 
\begin{equation}
F_{\mu\nu} := F_{\ \mu\nu}^{a}\partial_a = [D_{\mu}, D_{\nu}]  = (D_{\mu}\hat e_{\nu}^{\ a} - D_{\nu}\hat e_{\mu}^{\ a})\partial_a  \approx \, \frac{q}{m} \mathcal{F}_{\mu\nu} u^a\partial_a  \simeq -\mathrm{i}q\mathcal{F}_{\mu\nu},
\end{equation}
where $\mathcal{F}_{\mu\nu} := \mathrm{i}q^{-1}[\nabla_{\mu},\nabla_{\nu}] $. From now on, we will refer to $\mathbf{A}$ as the \textit{Colored Gravity Field}. Moreover, the torsion scalar yields the Yang--Mills scalar 
\begin{equation}
F^a_{\ \mu\nu} \simeq \frac{q}{m} \mathcal{F}_{\mu\nu}  u^a  \hspace{3mm} \mapsto \hspace{3mm}  F^a_{\ \mu\nu}F_a^{\ \mu\nu} \simeq \frac{q^2}{m^2}\mathcal{F}_{\mu\nu}\mathcal{F}^{\mu\nu}\, ,
\end{equation} 
which is valid for the spinor $\Psi$.


By introducing the perturbed co-tetrads in the \textit{torsion scalar} of the TEGR Lagrangian density (Eq. \ref{eq:lagrangian_TEGR}), and by including a source Lagrangian with energy density $\rho$, i.e., $\mathcal{L}_m = - \hat e \, \rho$, we arrive at \begin{equation}
\mathcal{L}  = \frac{\hat e}{2\kappa}\left(-\frac{1}{4}F^a_{\ \mu\nu} F_{a}^{\ \mu \nu} \right) + \mathcal{L}_m  \;\; \simeq \;\;  \frac{\hat e}{2\kappa}\left(-\frac{1}{4} \frac{q^2}{m^2}\mathcal{F}_{\mu\nu}\mathcal{F}^{\mu\nu} \right) + \mathcal{L}_m \, .
\end{equation} For a perfect fluid, the source energy is $\hat m = I_w m$ where $I_w := (1+3w)/(1+w)$, as shown in Eq. \ref{eq:lagrangian1b} of the Appendix \ref{sec:appendix}. Therefore, the total energy density is $\hat \rho = \bar \Psi I_w m  \Psi$, while the proper energy density is $\rho = \bar \Psi m  \Psi =  I_w^{-1} \hat \rho $. However, in our case the source energy is the gauge field $\mathbf{A}$.

Let $\mathbf{A}(0)$ be the gauge field evaluated at the position of the test particle, with four-current $\mathbf{q} = q\gamma_\mu$ and scalar charge $q$. The total density energy is $\mathbf{q} \bigcdot \mathbf{A}(0)$, where $\mathbf{A}(0) \approx\mathbf{q}/4\pi r_m$ is the potential in the particle, which is valid for short distances $r_m$ (see, for instance, \cite{Safarik2003}). Since our gauge fields behave as perfect fluids with $w = 1$ (cf. Eq. \ref{eq:metric_perturbation} of the Appendix \ref{sec:appendix}), the proper energy density has a factor $I_{w}^{-1} = 1/2$. Moreover, we need to introduce a normalized factor, $f_m = \mathrm{i}\gamma^\mu D_\mu/m + \mathcal{O}$, where $\mathcal{O}$ are the other terms related to either the Dirac equation or the Bargmann--Wigner equations, i.e., 
\begin{equation}
\rho \, = \, \bar \Psi \hat m_q f_m \Psi \; \; = \;\; \bar \Psi \; \frac{1}{2} \mathbf{q} \bigcdot \Delta \mathbf{A} \left(\frac{\mathrm{i}}{m}\gamma^\mu D_\mu  + \mathcal{O}\right) \, \Psi \, ,
\end{equation} 
where $ \Delta \mathbf{A} := \mathbf{A} - \mathbf{A}(0) \approx -\mathbf{A}(0)$ and $\bar \Psi := \Psi^{\dagger}\gamma^0$ is the Dirac adjoint, with $\Psi^{\dagger}$ being the Hermitian adjoint. On the macroscopic scale, with a large number of particles, the factor $f_m$ should be replaced by $f_m \mapsto \mathbf{1}$. Similarly, by changing $\bar \Psi  \hat m_q \Psi\,(\mathbf{x}) \mapsto \hat m_q \delta^4(\mathbf{x}-\mathbf{y})$, we obtain a density of the electric energy $\hat m_q$ at the point $\mathbf{y}$. In addition, the divergence term of the Ricci scalar, $-\partial_ \mu(T^{\nu\mu}_{\ \ \nu} \hat e /\, \kappa)$, is useful for the Einstein--Hilbert Lagrangian of GR to recover the classical Einstein field equations and the Maxwell's equations by considering the matter Lagrangian, $\mathcal{L}_m = -\hat e \rho \, \approx \, \frac{1}{2} \hat e \, \mathbf{q}\bigcdot\mathbf{A}(0)$ (see Sect. \ref{sec:firstorder}).

By hypothesis, we assume that the minimum distance corresponds to the particle event horizon $r_m = 2\mathrm{G}m$. This implies that the potential origin is $\mathbf{A}(0) \approx \mathbf{q}/(8\pi \mathrm{G} m) = \mathbf{q}/(\kappa\,m)$, thus yielding 
\begin{eqnarray} \nonumber
\mathcal{L} &\displaystyle = & \frac{\hat e}{2\kappa}\left(-\frac{1}{4}F^a_{\ \mu\nu} F_{a}^{\ \mu \nu} \right) \;\; - \;\;  \hat e \rho \; \simeq  \\ \nonumber
&  \simeq & \, \frac{\hat e}{2\kappa}\left(-\frac{1}{4}\frac{q^2}{m^2}\mathcal{F}_{\mu\nu}\mathcal{F}^{\mu\nu} \right)  \; + \;\frac{q^2 \hat{e}}{2\kappa\,m}  \, \bar \Psi \,  \left(\frac{\mathrm{i}}{m}\gamma^\mu D_\mu + \mathcal{O}\right)\,\Psi \;  \simeq  \\ \label{eq:lagrangian1}
{} & \simeq &  \frac{\hat e}{2\kappa}\frac{q^2}{m^2} \left(-\frac{1}{4} \mathcal{F}_{\mu\nu}\mathcal{F}^{\mu\nu}  + \bar \Psi \, \left(\mathrm{i}\gamma^\mu \nabla_\mu + m\mathcal{O}\right) \, \Psi \right) \, ,\end{eqnarray} which produces the same Euler--Lagrange equations than the Yang--Mills Lagrangian, ${\mathcal {L}}_{\mathcal{F}+\Psi}$ (recall Eq. \ref{lagYM}). However, this one has a different coupling factor between fermions and gravity compared to the first-order tetradic Palatini action, whose Lagrangian density is
\begin{equation}
{\mathcal {L}} = \hat e \, \left(\frac{1}{2\kappa} \hat e^{\mu }_{\ a}\hat e^{\nu }_{\ b}{R_{\mu \nu }}^{ab} +  \bar\Psi (i{\gamma}^{\mu }D_{\mu }-m) \Psi \right)\, , 
\end{equation} 
where $\hat e:=\det {\hat e_{\mu }}^{\ a} = \sqrt {-g}$, $\kappa = 8\pi\mathrm{G}$ and ${R_{\mu \nu }}^{ab}$ is the curvature of the spin connection. The interpretation of the different coupling factors is that the SU($1, 3$) Yang--Mills interactions would be particular cases of the colored gravity in which the spinors are coupled. In other words, due to the spatial gauge choice for the potentials, the gravitational constant, $\kappa$, does not affect in the first order of the perturbative interactions. 



\section{Electromagnetic limit and beyond}
\label{sec:geometric}

\subsection{Torsion-free components}

Before analyzing the classic limit, it is useful to check some properties of the Lorentz connection, since particle trajectories can be described by both the TEGR (with torsion) and the GR (without torsion) geodesics of the \textit{spacetime-particle} field. In fact, TEGR contorsion can be absorbed as a Riemann curvature effect \cite{Arcos2004, Krssak2019}. Similarly, electrodynamics can be modeled by torsionless Riemannian geometry because the gauge group is Abelian. This means that the subgroup $U(1) \subset SU(1, 3)$ will be considered in the following sections to describe the gauge potential $A$ (perturbation of the metric). The perturbed co-tetrads lead to a perturbed spin connection, whose (torsionless) components $\omega_\mu^{\ ab}$ are (Eq. \ref{connection}):
\begin{eqnarray}  \nonumber 
\omega_{\mu }^{\ ab}= \hat e_{\nu }^{\ a} \overset{\circ} D \hat e^{\nu b} = \hat e_{\nu }^{\ a}(\partial _{\mu }\hat e^{\nu b}+\overset{\circ}\Gamma{}_{\ \sigma \mu }^{\nu } \hat e^{\sigma b}) \\ \nonumber
\hspace{3mm} = {\textstyle \frac {1}{2}}(\partial _{\mu }\hat e_{\nu }^{\ b}-\partial _{\nu } \hat e_{\mu }^{\ b})-{\textstyle \frac {1}{2}}\hat e^{\nu b}(\partial _{\mu } \hat e_{\nu }^{\ a}-\partial_{\nu }\hat e_{\mu}^{\ a})-{\textstyle \frac {1}{2}} \hat e^{\rho a}\hat e^{\sigma b}(\partial _{\rho }\hat e_{\sigma c}-\partial_{\sigma}\hat e_{\rho c})\hat e_{\mu }^{\ c} 
\\ \nonumber
\hspace{6mm} \approx {\textstyle  \frac {1}{2}}( \partial _{\mu }\hat e^{a b} 
- \partial _{\mu }\hat e^{b a} 
-\partial^a \hat e_{\mu }^{\ b}  
+ \partial^{b}\hat e_{\mu }^{\ a} - \eta^{\rho a} \eta^{\sigma b} \eta_{c d} (
\partial^{a} \hat e_\sigma^{\ d}
-\partial^{b} \hat e_\rho^{\ d})\delta^c_\mu) \;\; 
\\ \label{connection1}
\hspace{3mm} \approx -(\partial^a \phi^{\ b}_{\mu} -  \partial^{b} \phi^{\ a}_{\mu}) 
\approx - {\textstyle \frac{q}{m}} \left( \partial^a(  A^b u_{\mu}) - \partial^b(  A^a u_{\mu}) \right). 
\end{eqnarray} 
Two interesting properties of the connection are found for the application to spinors $\Psi$ and for contractions with $\gamma^\mu$ (useful for vector fields). Taking into account the Lorentz generators for spinors, $ S_{a b}^\Psi = i[\gamma_a, \gamma_b]/4$, we obtain
\begin{equation}
\begin{split}
\omega_{\mu }^{\ ab} S_{a b}^\Psi & \approx { \frac{\mathrm{i}}{4}}\omega_{\mu }^{\ ab} [\gamma_a, \gamma_b] \\ \label{eq:connection2}
\hspace{6mm}& \approx -  { \frac{\mathrm{i}}{4}} {\textstyle \frac{q}{m}}  \left( \partial^a( A^b u_{\mu}) - \partial^b(  A^a  u_{\mu}) \right) (\gamma_a\gamma_b - \gamma_b\gamma_a)
\approx  0,
\end{split}
\end{equation}
which means that the torsion-free connection does not provide additional terms to the total covariant derivative, and it requires to take into account derivatives of both the spin connection and the gauge covariant derivative, e.g., $\hat D_{\mu} =\hat\partial_{\mu} - \mathrm{i}\omega_{\mu}^{\ ab} S_{ab}/2 = \hat\partial_\mu + 0 \simeq \nabla_\mu$. Alternatively, one can incorporate the contorsion part of the Lorentz connection, thus obtaining in a natural way the total covariant derivative $D_\mu \simeq \nabla_\mu$.

 On the other hand, the connection form for vector fields is obtained by the contraction with $u^{\mu}$, and it is also useful to obtain the contraction with $dx^\mu$
 \begin{eqnarray}
\label{eq:connection3}
\omega_{\mu }^{\ ab} u^{\mu} \approx - {\frac{q}{m}}  \left( \partial^a A^b - \partial^b A^a \right) \approx - { \frac{q}{m}} \mathcal{F}^{ab}, \\
\omega^{ab} := \omega_{\mu }^{\ ab} dx^{\mu} \approx \partial^b \phi^{a} -  \partial^{a} \phi^{b}  = - \frac{q}{m}\mathcal{F}^{ab} d\tau,
\end{eqnarray}
where the equality $\gamma_{\mu}\gamma^{\mu} = \gamma^{\mu}\gamma_{\mu} = 4$ has been used and, in particular, $\gamma^{\mu}(\partial^a \gamma_{\mu}) = 0$. Moreover, recall that $\mathcal{F}^{a b} := {\partial}^{a}A^{b} -{\partial}^b A^{a}$ are the components of the U(1) field-strength tensor, that is, the components of the 2-form Faraday curvature $\mathbf{F} := {\frac {1}{2}}\mathcal{F}_{\mu \nu }\mathrm {d} x^{\mu }\wedge \mathrm {d} x^{\nu } =\mathrm {d} A =(\partial _{\mu }A_{\nu })\mathrm {d} x^{\mu }\wedge \mathrm {d} x^{\nu }$. Notice that details on possible non-abelian effects of the sources are lost if the contorsion is not added. 

For a vector field $v^a$, the total time derivative $\mathbf{D_\tau} := \hat \gamma^\mu D_\mu = \gamma^\mu \partial_\mu$ behaves well. Recall that the equation of motion for a free particle is given by $\mathbf{D_\tau} v^a = 0$ (Eq. \ref{totaltime}), that is, \begin{equation} 
\mathbf{D_\tau}  v^a := \hat \gamma^\mu D_{\mu} v^a =  \mathbf{d_\tau} v^a  + \eta_{bc} \omega_{\mu}^{\ ac}  {u}^{\mu} v^b = 0 \hspace{4mm} \Rightarrow \hspace{4mm}  \mathbf{d_\tau} v^a \approx \frac{q}{m}  \mathcal{F}^{a c} v_{c}, \;\;
\end{equation} which is the equation of a particle subject to a Lorentz force. As mentioned in the previous sections, the acceleration of spinless particles can be modelled by both the torsion tensor (like an \textit{external} force) and the geodesic equation within the pseudo-Riemannian geometry. That is to say, the description of the movement is provided by the Ricci coefficient of rotation and, therefore, it is also into the Ricci tensor. 

In local coordinates, since the components of $\mathrm{d}\mathcal{F}^{a}_{\ b}$ are sixteen real functions, the Lorentz curvature 2-form components are
\begin{equation}  R_{\;\,b}^{a}  = \mathrm{d}\omega _{\;\,b}^{a}+\omega _{\;c}^{a}\wedge \omega _{\;\,b}^{c}
\approx - \frac{q}{m}\mathrm{d}\mathcal{F}^{a}_{\ b}  d\tau + \left(\frac{q}{m}\right)^2 \mathcal{F}^a_{\ c} \wedge\mathcal{F}^c_{\ b}  {d\tau}^2.
\end{equation}
Similarly, applying $\hat e_\mu^{\ a}=e_\mu^{\ a}+\phi_\mu^{\ a}$ adequately to the Ricci curvature, taking into account Eq. \ref{connection1}, the first-order Ricci scalar for the Hilbert--Einstein action is 
\begin{eqnarray} \nonumber
R & := & R_{ab} \wedge \star \hat e^{ab} = R_{\mu\nu a b}\hat e^{\mu\nu ab} = {{{\frac{1}{2}}(\partial_\mu \omega_{\nu ab} + \omega_{\mu a}^{\ \ c}\omega_{\nu cb})}}\left( \hat e^{\mu a} \hat e^{\nu b} - \hat e^{\mu b} \hat e^{\nu a}\right) \approx 
\\ \nonumber
& \approx & - {\textstyle \frac{1}{2}} {\textstyle \frac{q}{m}}  \partial_\mu \left( \partial_a(  A_b u_{\nu}) - \partial_b(  A_a u_{\nu}) \right) \left( \hat e^{\mu a} \hat e^{\nu b} - \hat e^{\mu b} \hat e^{\nu a}\right)  + 0 \approx \\
& \approx & - {\textstyle \frac{q}{m}}  \partial^a \left( \partial_a(  A_b u^b) - \partial_b(  A_a u^b) \right)    \approx  - {\textstyle \frac{1}{m}}  \partial^a \partial_a(  A_b j^b)   
\end{eqnarray} 
where $\partial^a \partial_b(  A_a u^b) = 0$ and $j^b := q u^b$ are considered. This curvature scalar density is consistent with the electromagnetism prescription, as shown in the following subsection.

\subsection{Physical prescription}
\label{sec:Hypotheses}

According to the above developments, the electrodynamics equations are obtained by the choice of a potential energy origin, which causes curved geodesics. This subsection is focused on the description of that idea and its consequences in a classical way, and is supported by auxiliary calculations in the Appendix \ref{sec:appendix}. Of course, electromagnetism is defined by the Abelian subalgebra $\mathfrak{u}(1) \subset \mathfrak{su}(1, 3)$. Therefore, the metric perturbation $\phi^{\ a}_\mu = A^a u_\mu$ is now real, isomorphic to the classical linear perturbation in gravity.

\textit{Spacetime-particle geodesics}. For classical gravitational interactions, the geodesics generated by a source energy do not depend on the energy (inertial mass) of the point object in ``free fall''. However, the metric perturbed by a gauge potential now differs from that situation, since it depends on the mass $m$ and the charge $q$ of every ``test particle'', forming a new indivisible entity of \textit{spacetime-particle field}. In other words, the interaction energy ($qA$) is not the same than the inertial energy ($m$). In order to ensure continuity in the metric tensor field, $g$, it is required that all the quantities are also fields (constant or not), that is, \begin{equation} \label{eq:metricc} g_{\mu \nu} (x)  \approx \eta_{\mu \nu} + 2\phi_{\mu \nu}(x)  \approx  \eta_{\mu \nu} + 2\frac{q(x)}{m(x)} {A}_{\mu}(x) u_{\nu}(x).
\end{equation} where the fields $m$ and $q$ are not constant due to the dependence on the position of the particles. The potential vector field, $A(x)$, needs to be the sum of all of the Lienard-Wiechert-like potentials evaluated at $x$, including the gauge field generated by the quantum numbers ($q(x)$) and any possible external fields. 

\textit{Two-particle interaction}. Consider a system with only two (almost) point particles, `1' and `2'. For particle `1', let $q_1$ be its electric charge, $u_1^\alpha$ be its 4-velocity, $y=(t,\vec r_1)\in\mathbb R^4$ its location and $A_1$ be the vector field caused by it. Analogously, for the particle `2', let $q_2$ be its electric charge, $u_2^\alpha$ be its 4-velocity, $m$ be its mass, $x=(t,\vec r_2)$ be its location and $A_2$ be the vector field caused by it, which is assumed to be negligible, i.e., $|A_2| \ll |A_1|$. Therefore, the total vector field, $A(x) \approx A_1(x)$, is essentially external and the metric evaluated at $x$ should be interpreted, on a macroscopic scale, in one of the following two ways,\begin{eqnarray} \label{eq:metric3a} 
\hspace{14mm} g_{\mu \nu}^{\,\,(a)}  \approx \eta_{\mu \nu} + 2\phi_{\mu \nu}^{\,\,(a)}  \approx \eta_{\mu \nu} + 2\frac{q_2}{m} {A}_{1}^{(2)} u_{2\mu} u_{2\nu}, \\  \label{eq:metric3b}
\hspace{14mm} g_{\mu \nu}^{\,\,(b)}  \approx \eta_{\mu \nu} + 2\phi_{\mu \nu}^{\,\,(b)} \approx \eta_{\mu \nu}  + 2\frac{q_2}{m} {A}_{1(\mu} u_{2\nu)}m,
\end{eqnarray} 
where $A_{1}^{(2)} \approx A_{1\gamma} u_{2}^{\gamma}$ and ${A}_{1(\mu} u_{2\nu)}  := ({A}_{1\mu}u_{2\nu} + u_{2\mu}{A}_{1\nu})/2 = A_1({u}_{1\mu}u_{2\nu} + u_{2\mu}{u}_{1\nu})/2$ are defined by a scalar function $\zeta(r)$ such that $A_{1\gamma}(r) = q_1u_{1\gamma}\zeta(r)$, and the relative distance $r := \max\{|(x^\mu-y^\mu) u_\mu|, r_m\}$ is measured with respect to a minimum value $r_m$. By defining the four-currents as $J_{1\nu} := q_1 u_{1\nu}$ and $J_{1\nu} := q_2 u_{2\nu}$, we can identify the semi-classical perturbation $\Phi_{\alpha\beta}$:
{\small
\begin{eqnarray}
g_{\mu \nu}^{\,\,(a)} &\approx &\eta_{\mu \nu} + \frac{2}{m} \zeta(r){J}_1^{\gamma} J_{2\gamma} u_{2\mu}u_{2\nu}  \approx: \eta_{\mu \nu} + \Phi_{\mu \nu}^{\,\,(a)} \hspace{0.5mm} \Rightarrow \hspace{0.5mm}  \Phi_{\mu \nu}^{\,\,(a)} = \frac{4\mathrm{G}}{r_m} \zeta(r){J}_1^{\gamma} J_{2\gamma} u_{2\mu}u_{2\nu}, \; \label{eq:metrica1}\\
g_{\mu \nu}^{\,\,(b)} &\approx& \eta_{\mu \nu} + \frac{2}{m} \zeta(r){J}_{1(\mu} J_{2\nu)}  \approx: \eta_{\mu \nu} + \Phi_{\mu \nu}^{\,\,(b)} \hspace{7mm} \Rightarrow \hspace{3mm}  \Phi_{\mu \nu}^{\,\,(b)} = \frac{4\mathrm{G}}{r_m} \zeta(r){J}_{(1\mu} J_{2\nu)},\label{eq:metrica2}
\end{eqnarray}
}
where the minimum distance (or spatial gauge) has been chosen in the \textit{event horizon} of the test particle `2', i.e., $r_m := 2\mathrm{G}m$ with $w=1$ (see Eqs. \ref{eq:metric_perturbation} and Eq. \ref{eq:perturbed_energy} of Appendix \ref{sec:appendix}). Taking into account the relative distance $r$, the scalar function is $\zeta(r) = \mathrm{k}/r$, where $\mathrm{k} = 1/(4\pi)$ is the Coulomb constant with permittivity $\epsilon_0 \equiv 1$ in natural units. Now, one can identify a Lienard-Wiechert-like potential $\Phi$ from the pertubed metric (Eq.~\ref{eq:metrica2}) 
 such as
\begin{equation}\label{eq:metricb}
\begin{split}
\Phi_{\mu \nu}^{\,\,(b)}  \approx\;  & 4\frac{\mathrm{G}}{\mathrm{k}} \zeta_m {A}_{(1\mu} J_{2\nu)} \approx 4{\mathrm{G}}\,\zeta_m \frac{{J}_{(1\mu} J_{2\nu)}}{r}\approx 4\frac{\mathrm{G}}{\mathrm{k}} \frac{\mathrm{k}q_2}{r_m} \frac{kq_1}{r}{u}_{(1\mu} u_{2\nu)}\\
 & \approx 4\frac{\mathrm{G}}{\mathrm{k}} \, {A}_{(1\mu} \, A_{2\nu)},   
\end{split}
\end{equation}
where $\zeta_m := \zeta(r_m) = \mathrm{k}/{(2\mathrm{G}m)} = 1/(8\pi\mathrm{G}m) = 1/(\kappa m)$, while $A_1 := A_1(r)$ and $A_2 := A_2(r_m) := J_2\zeta_m$, as the reference is at the particle `2'.

\subsection{Lorentz force and Maxwell equations}

\textit{Lorentz force}. The electric energy perturbation leads to the Lorentz force. To prove this, it is enough to use linearized gravity (Eq. \ref{eq:lagrangian1b} of Appendix \ref{sec:appendix}) and the metric of Eq.~\ref{eq:metric3a} or Eq.~\ref{eq:metric3b}. For both metrics, the Lagrangian functional for a test particle `2' is the same:
\begin{equation} \label{eq:lagrangian2}   
L \hspace{1mm} = \hspace{1mm} \frac{m}{2}g_{\alpha\beta}u_2^{\alpha}u_2^{\beta}
\hspace{1mm} \approx \hspace{1mm}  \frac{m}{2}\eta_{\alpha\beta}u_2^{\alpha}u_2^{\beta} + A_1^{\gamma} q_2 u_{2\gamma}.
\end{equation}
The second-order term of the metric is omitted because its effects are negligible for point particles (see Appendix \ref{appendix:second-order-lorentz}). By defining the Faraday tensor as $F_{\gamma\alpha} := \partial_{\gamma}A_{1\alpha} - \partial_{\alpha}A_{1\gamma}$, the Euler--Lagrange equations (Eq. \ref{eq:Euler.lagrange0} in Appendix \ref{sec:appendix}) are
\begin{equation}
\begin{split}
\frac{d}{d\tau}\left(m u_{2\alpha} + q_2 A_{1\alpha} \right) - q_2u_2^{\gamma}\frac{\partial  A_{1\gamma}}{\partial x_2^{\alpha}}=& \frac{d}{d\tau}\left(m u_{2\alpha}\right) + q_2u_2^{\gamma} \frac{\partial A_{1\alpha}}{\partial x_2^{\gamma}}  - q_2u_2^{\gamma}\frac{\partial  A_{1\gamma}}{\partial x_2^{\alpha}}\\
=& \frac{d}{d\tau}\left(m u_{2\alpha}\right) + q_2u_2^{\gamma}\mathcal{F}_{\gamma\alpha}=0,
\end{split}
\end{equation}

which is the Lorentz force.

\textit{Perturbation sources}. The \textit{Source Analysis} (SA) is defined by considering that the total potential field $A(x)$ is generated only by the source `1' in $x = x_1$, that is, $q(x) = q_1(x)$, $J(x) = J_1(x)$ and $A(x) = A_1(x)$. Nevertheless, classically it may be more interesting to consider the \textit{Interaction Analysis} (IA), with a test particle `2' with $q_2 \ll q_1$ that is moving parallel to `$1$' (i.e., $u := u_1 = u_2$) to reproduce a stationary frame. For both cases, the metric of Eq.~\ref{eq:metrica1} is equal to Eq. \ref{eq:metrica2}, and the IA recovers the SA when $q_2$ is replaced by $q_1$. For instance, consider the following field:
\begin{equation*}
\Phi_{\mu \nu}\equiv\Phi_{\mu \nu}^{\,\,(b)}  \approx 4\frac{\mathrm{G}}{\mathrm{k}} \zeta_m {A}_{1\mu} J_{2\nu} = 4\frac{\mathrm{G}}{\mathrm{k}} \zeta_m {A}_{1\mu} q_{2} u_\nu.
\end{equation*}
Using the first-order Einstein field equations (Eq.~\ref{eq:ricci1} of Appendix \ref{sec:appendix}), we find that 
\begin{equation}\label{eq:einstein}
\begin{split}
 R^{(1)}_{\mu\nu} \approx 8\pi \left(P_{\mu \nu} -\frac{P}{2}g_{\mu \nu} \right) &\Longrightarrow \hspace{3mm}
 - \frac{1}{2} \partial^\gamma\partial_\gamma \Phi_{\mu \nu}  \approx 8\pi \mathrm{G} \hat{\rho}_q u_\mu u_\nu\\
 & \Longrightarrow \hspace{3mm}
-2\frac{\mathrm{G}}{\mathrm{k}} \zeta_mq_{2} \partial^\gamma\partial_\gamma ({A}_{1\mu} u_\nu)  \approx 8\pi \mathrm{G} \hat{\rho}_q u_\mu u_\nu,
\end{split}
\end{equation}
where $w=1/3$ is required, which corresponds to a perfect fluid, and the total energy density (including pressure effects and omitting retarded terms) equals to 
\begin{equation} \label{eq:important}
\hat{\rho}_q \approx - \frac{\mathrm{d}(J_1^{\gamma}J_{2\gamma}\zeta_{m})}{\mathrm{d}V} \approx 
-q_2 Q\,\zeta_m, 
\end{equation} 
where the (rest) charge density is given by
\begin{equation*}
Q(y)\, := \frac{\mathrm{d}q_1}{\mathrm{d}V} :\approx q_1 \int\frac{\delta^4(y-x)}{\sqrt{-g}}\mathrm{d}\tau.    
\end{equation*}
Since the test particle appears on both sides of Eq.~\ref{eq:einstein}, the dependence on $\zeta_m q_2$ vanishes. As expected, considering the SA instead of the IA, it also vanishes. With this, the Maxwell equations are obtained in Lorentz gauge formulation:
\begin{equation}
\partial^\gamma\partial_\gamma A_{1\mu} \approx 4\pi k j_{\mu},
\end{equation}  
with $j_{\mu} := Q\,u_{\mu}$. Note that the harmonic coordinate condition $\partial^{\alpha} \Phi_{\alpha \beta} = \frac{1}{2} \eta^{\alpha\gamma}  \partial_{\beta} \Phi_{\alpha \gamma}$ for $\Phi_{\alpha\beta} \approx q_2\zeta_m A u_{\alpha}u_{\beta}$ is equivalent to the Lorentz gauge ($\partial^{\gamma} A_{\gamma} = 0$) when the four-velocity $u^{\beta}$ is applied to both sides of the Eq.~\ref{eq:einstein}, as  $A$ is an electrostatic field ($u^{\beta}\partial_{\beta} A = 0$) and $0=\partial^\gamma (u^\beta u_\beta) = 2 u^\beta  \partial^\gamma u_\beta$. To complete this result, the homogeneous equation $\nabla_{[\alpha }\mathcal{F}_{\beta \gamma ]}= \nabla_{[\alpha }\nabla A_{[\beta \gamma] ]} = 0$ is easily obtained from the first Bianchi identity, $R^{\sigma}_{\ [\beta \gamma \delta]}=0$, by applying the identity $A_{\sigma}R^{\sigma}_{\ \beta \gamma \delta}= \nabla_{[\gamma} \nabla_{\delta]} A_\beta$. This is also trivially found by using the 2-form representation of the Faraday tensor ($\mathbf{F} = \frac{1}{2}\mathcal{F}_{\mu\nu} dx^\mu \wedge dx^\nu = \mathrm{d}A$), i.e., $\mathrm{d}\mathbf{F} = 0$, since the second exterior derivative is zero ($\mathrm{d}^2 \equiv 0$).

\subsection{First-order Lagrangian density}\label{sec:firstorder}

The Lagrangian density of matter ~\cite{Harko2010} is $\mathcal{L}_M = -\rho {\sqrt {-g}}$, where $\rho = I_w^{-1}\hat \rho$ is the proper energy, which is related to the effective total energy density. For electromagnetism, one has $\hat \rho \approx \hat \rho_q = -j_{1\mu}J_2^{\mu}\zeta_m$, where $j_{1\mu} := Qu_{1\mu}$ is the density of the source four-current and $J_{2\mu} := q_2u_{2\mu}$ is a test-particle 4-current. With the first-order Ricci tensor, $R^{(1)}_{\mu\nu} = -\frac{1}{2}\partial_{\gamma} \partial^{\gamma} \Phi_{\mu\nu}$, and Eq.~\ref{eq:metricb}, the Einstein--Hilbert Lagrangian density is 
\begin{equation}
\begin{split}
 {\mathcal {L}}[g^{\mu\nu}] = &{\sqrt {-g}}\left(\frac{g^{\mu\nu}{R_{\mu \nu }}}{16\pi\mathrm{G}}  - \rho \right)\\
 \approx & { \sqrt {-g}}\,{g^{\mu\nu}} \left(\frac{-1}{8\pi\mathrm{k}}  \partial_{\gamma}\partial^{\gamma}(\zeta_m A_{(1\mu}J_{2\nu)}) + \frac{1}{2} j_{(1\mu}J_{2\nu)}\zeta_m \right).
\end{split}
\end{equation}

The Euler--Lagrange equations depend only on the fields $g^{\mu\nu}$. Setting the latter equal to zero and contracting with respect to the velocity of the test particle $u_2^\nu$, we conclude that 
\begin{equation*}
\frac{1}{4\pi\mathrm{k}}  \partial_{\gamma}\partial^{\gamma}(A_{(1\mu}J_{2\nu)}) = j_{(1\mu}J_{2\nu)} \;\; \Longrightarrow  \;\;  \partial_{\gamma}\partial^{\gamma}A_{1\mu} = 4\pi\mathrm{k}j_{1\mu},
\end{equation*}
as expected. Observe that electromagnetism is formulated as a particular case of \textit{colored gravity}.

\subsection{Charge-energy and source-gravity relationships}

The integration of Eq. \ref{eq:important} leads to an electrical energy potential with origin in $r_m$, which is calculated by the displacement of a test particle `2' between $r_m$ and $r$, both in a stationary reference system. In other words, the electrical energy suffered by `2' (since we take $r_m$ in `2') is  \begin{equation}
\label{eq:electric.energy} 
m_{q}(r) :\approx J_1^{\alpha}J_{2\alpha}\Delta\zeta(r) + O(\partial\zeta^2) =  J_1^{\alpha} \Delta A_{2\alpha}(r) + O((\partial A_2)^2),
\end{equation}where $\Delta A_2^{\alpha}(r) := J_2^{\alpha}(\zeta(r) - \zeta(r_m))$ is the hypothetical electromagnetic 4-potential of `2'. The possible second order term, $O((\partial A_2)^2)$, is an electromagnetic self-energy of the particle `2' (e.g., from the Faraday tensor, $F^{\mu\alpha}F^{\nu}{}_{\alpha} - \frac{1}{4} \eta^{\mu\nu}F_{\alpha\beta}F^{\alpha\beta}$). By comparing to the limit case of the semi-classical gravity perturbation with $w=1$, $\Phi_{\alpha\beta}^{(a)} = 4\phi(r)u_{\alpha}u_{\alpha}$, it \textcolor{black}{is} easy to identify an approach for the gravitational potential: 
\begin{equation} \label{eq:einstein1.sol0}   
\phi(r) \approx \frac{\mathrm{G}J_1^{\gamma}J_{2\gamma}\zeta_{m}}{r} = \frac{\mathrm{G}}{\mathrm{k}} {A}_{1\mu} J_2^{\mu} \zeta_m =  \frac{\mathrm{G}}{\mathrm{k}} A_{\mu}(r)J^{\mu}(r) \zeta_m,
\end{equation} where $A_{\mu}(r)$ and $J^{\mu}(r)$ are the total potential field vector and the 4-current, respectively, evaluated at $r$. In the SA case, it is found that $\phi(r) = r_q^2/(r_m r)$, where $r_q^2 := \mathrm{G\,k}q^2$ is the characteristic length scale of the  Reissner--Nordstr\"om metric, whose electromagnetic term is $\phi(r) = r_q^2/r^2$ \cite{Giorgi2019}. To make compatible both metrics, it is necessary to use the (Kerr--Schild) double-copy solution, i.e.,

\begin{equation} \nonumber
\Phi_{\mu\nu}(r) = 4\phi(r)u_\mu u_\nu \sim - 2\frac{\mathrm{G}}{\mathrm{k}} \hat A_\mu \hat A_\nu = -\kappa \hat A_\mu \hat A_\nu,
\end{equation}
where $\hat A_\mu := \Delta A_\mu = A_\mu (r) - A_\mu(r_m)$. Note that the residual constant is omitted.

Consistently with linearized gravity (Eq. \ref{eq:perturbed_energy} of Appendix \ref{sec:appendix}), the electric-based perturbation energy suffered by `2' is also $m_{q} \approx {m}I_w \Delta\phi(r) = {m}I_w (\phi(r)-\phi(r_{m}))$ when $\phi(r)$ is predominantly of electrical origin (Eq.~\ref{eq:einstein1.sol0} and Eq.~\ref{eq:perturbed_energy} of Appendix \ref{sec:appendix}). Setting it equal to Eq. \ref{eq:electric.energy}, one can easily check that 
\begin{equation} \label{eq:perturbation} m_{q}(r) \hspace{2mm} \approx \hspace{2mm} J_1^{\alpha} \Delta A_{2\alpha}(r)  \hspace{2mm} \approx \hspace{2mm}  {m}I_w (\phi(r)-\phi(r_{m})).
\end{equation}
Since the origin is in the test particle ($r_m = 2\mathrm{G}{m}$), the total potential energy ($r \to \infty$) of Eq.~\ref{eq:perturbation} represents the total source energy with $w=1/3$, that is, $m_q(\infty) = -J_1^{\alpha}J_{2\alpha}\zeta_m = -q_1q_2 \zeta_m = - q_1q_2/(\kappa m) < 0$. For example, assuming that the self-energy of a $q$-charged particle is $m_q(\infty) = m$, it follows the Planck mass-charge relationship, \begin{equation}
    \kappa\,m^2 = - q^2\,. 
\end{equation} Therefore, in order to translate some semi-classical results between gravity and electromagnetism, it is useful to define the application: 
\begin{eqnarray} \label{eq:transformation}
\pazocal{T} :  & \mathbb{R} \cdot \mathcal{Q} \hspace{2mm} \rightarrow    \mathrm{i}\,\mathbb{R} \\
&    C \cdot q \hspace{2mm} \mapsto    \pazocal{T}(C \cdot q) :=  C \cdot \pazocal{T}(q) :=  C \cdot \mathrm{i}\sqrt{\frac{2G}{\mathrm{k}}}\, m  = C \cdot \mathrm{i} \sqrt{\kappa} \, m,
\end{eqnarray}where $\mathcal{Q}$ is the space of the coupling constants, $\mathrm{k} = 1/4\pi$ and $\kappa = 8\pi\mathrm{G}$, while $q$ and $m$ are the charge and the mass, respectively, of a point particle in natural units ($1 = c = \hslash = \epsilon_0$). Thanks to this transformation, Eq.~\ref{eq:einstein1.sol0} recovers the classical gravitational field, $\phi(r) = - \mathrm{G}m/r < 0$, which is a (simple) \textit{gravity spacetime} perturbation. In fact, the equivalent energy perturbation for the SA case (with $J^{\gamma} := J_1^{\gamma} = J_2^{\gamma}$) is the equivalence $\pazocal{T}(m_q(\infty)) = \pazocal{T}(-q^2\zeta_m) = m$, and Eq.~\ref{eq:metrica1} leads to the linearized gravity with $w=1$. 

An important outcome is that the source energy is negative when the total integral is taken into account ($m_q(\infty) = -J^{\gamma}J_{\gamma}\zeta_m = -q^2\zeta_m < 0$ for the SA). To obtain positive energy (mass), it is necessary to combine electric charges with different signs and make quantum-electrodynamic corrections. The effective electrical neutrality of the matter sources is key in obtaining an effective positive energy.

The classical gravity interaction can be understood as a sum of the effects provided by the effective energy interactions of the Colored Gravity, which could include all the other interactions (source effects). As a drawback, the proposed semi-classical theory cannot predict fundamental parameters, for instance the Coulomb or fine-structure constants. However, the double-copy solution of the perturbed metric could provide relevant information on the fundamental interaction, and higher-order corrections can be analyzed in future work. The most important aspects to be addressed are: i) the quantization procedure to obtain a quantum field theory of gravity, while being compatible with the standard model, and ii) the extension of the model at a cosmological scale to explore early stages of the Universe. 

To extend the spacetime algebra to a cosmological scale, the proposed methodology can be applied to the hyperconical model of Monjo \cite{Monjo2017,MCS2020} by using either non-local or asymptotic symmetries \cite{Krasnikov2021, Fabbrichesi2021}. This is a key at cosmological scales because the hyperconical model explains phenomenology similar to that presumably due to \textit{dark quantities}, but it does not produce photons by itself since it establishes a `zero active' energy \cite{MCS2023, Monjo2024}.

\section{Conclusions}
\label{sec:discussion}

This work is based on perturbed tetrads and spacetime generators with a covariant Wilson-line term (in the Lorentz gauge) representing the modified ground state of the spacetime generators according to the observer's frame of reference.
 
The introduction of that small perturbation to the local coordinates formulates gauge interaction phenomena for SU(1, 3) Yang--Mills dynamics. In particular, unitary (gauge) transformations of local phases are equivalent to the change of reference system from local coordinates to the observer's reference frame. In fact, the expectation value of translation \textit{gauge-like} covariant derivative is set to coincide with the SU(1, 3)-gauge covariant derivative, and the transformation can be simplified as a change of the reference frame (like a choice of the gauge potential origin).

As a general output for SU(1, 3) interactions, the metric depends on the field features (coupling factors) at each position. To ensure continuity in the metric tensor field, the coupling factors should also be fields, and it is forbidden that two different features, e.g., two masses, are at the same 4-position (this is similar to the Pauli exclusion principle in fermionic statistics). In fact, SU(1, 3) can be formulated from complexified perturbations of the Clifford--Dirac algebra and, therefore, the results of this paper could provide basic ideas for colored gravity. 

As a particular limit, the classical electromagnetism with Lorentz force and Maxwell equations is recovered from the Einstein Field Equations by using the perturbed tetrad. On a macroscopic scale, it is equivalent to setting a spatial gauge, $r_m$, for the energy origin in the electric 4-potential field. This origin is set in the event horizon of each test particle with energy $m$ and the equation of state $w=1$, that is, $r_m = 2\mathrm{G}m$.

In this context, spacetime would present continuous rough fields given by the small gauge-based perturbations (linked to the coupling constants and the particle quantum numbers). Moreover, the torsion of the developed metric ($g \sim \eta + A \otimes A$) can be interpreted with the picture of a spacetime with a double-helix-like structure based on pairs $A \otimes A$ of entangled virtual $\mathfrak{su}(1,3)$ bosons (e.g., photons). These potential fields are the connection (force carrier or virtual-particle exchange) of a double-copy gauge transformation, generated by an extended Poincaré algebra $\mathbb{R}^{1,3} \oplus \mathfrak{u}(1,3)$.

In conclusion, future works should inspect further the double-copy gauge viewpoint of the metric, addressing quantization procedures and Lagrangian density corrections, as well as on the extension of the perturbed algebra on a cosmological scale. Work in both directions is currently in progress.



\backmatter

\bmhead{Acknowledgements} The authors greatly appreciate all the very useful comments and suggestions of Prof. Fernando Barbero, Prof. Luis Acedo, Prof. Piergiulio Tempesta and the anonymous referees. The second author (ARA) is supported by the Ministerio de Universidades, Spain, under an FPU grant and partially supported by the Ministerio de Ciencia e Innovación, Spain, under grant PID2021-126124NB-I00. The third author (RCS) acknowledges the
partial financial support from the research grant  PID2019-106802GB-I00/AEI/10.13039/501100011033 (AEI/ FEDER, UE). 

\section*{Declarations}

\begin{itemize}
\item \textbf{Author contribution}: The original idea, symbolic calculus and main drafting of the paper was performed by RM, while ARA and RCS revised the formal consistency of the definitions and results. All authors read and approved the final paper.
\item \textbf{Funding}: Ministerio de Ciencia e Innovación, Spain, under grant PID2021-126124NB-I00 and Research grant number PID2019-106802GB-I00/AEI/10.13039/501100011033 (AEI/ FEDER, UE)
\item  \textbf{Conflict of interest}: Not applicable
\item  \textbf{Ethics approval}: Not applicable
\item \textbf{Consent to participate}: Not applicable
\item \textbf{Consent for publication}: Not applicable
\item \textbf{Code availability}: Not applicable
\item \textbf{Materials availability}: Not applicable
\end{itemize}

\noindent
If any of the sections are not relevant to your manuscript, please include the heading and write `Not applicable' for that section. 

\bigskip

\begin{appendices}

\section{Linearized gravity and gravitomagnetism}\label{sec:appendix}

\subsection{Linearised gravity for point particles}
\label{sec:scalar} 

To support the analysis of the classical limit in a macroscopic scale, in this Appendix we recall some results of Classical Relativity, especially from the linearized formalism and the gravitomagnetism solution \cite{Capozziello2009, Skagerstam2019}. Using local coordinates with an affine connection, the Ricci curvature tensor components ($R_{\alpha \beta}$) are computed from the Riemann curvature tensor components ($R_{\alpha \beta \gamma}^{\mu}$), that is,
\begin{equation} \label{eq:Ricci0}   
R_{\alpha \beta} := R_{\alpha \mu \beta}^{\mu} = \partial_{\mu}\Gamma_{\alpha \beta}^{\mu} - \partial_{\beta}\Gamma_{\mu \alpha}^{\mu} + \Gamma_{\rho \mu}^{\mu}\Gamma_{\alpha \beta}^{\rho} - \Gamma_{\rho \beta}^{\mu}\Gamma_{\alpha \mu}^{\rho},
\end{equation} where $\Gamma_{\alpha \beta}^{\mu}$ are the Christoffel symbols of the second kind, which for the Levi--Civita connection $\hat\nabla$ satisfy $\hat\nabla g = 0$, such that $g$ is free of torsion. Therefore, for a perturbed metric, $g_{\alpha\beta} \approx \eta_{\alpha\beta} + \Phi_{\alpha\beta}$, the first-order Christoffel symbols are 
\begin{equation} \label{eq:Christoffel2}   
\Gamma_{\alpha \beta}^{\mu} = \frac{1}{2} g^{\mu \nu}\left(\partial_\alpha g_{\nu \beta} + \partial_\beta g_{\alpha \nu} - \partial_\nu g_{\alpha \beta} \right) \approx
\frac{1}{2} \eta^{\mu \nu}\left(\partial_\alpha \Phi_{\nu \beta} + \partial_\beta \Phi_{\alpha \nu} - \partial_\nu \Phi_{\alpha \beta} \right).
\end{equation} 
By using the harmonic coordinates condition, i.e., $\eta^{\beta\gamma} \partial_{\gamma} \Phi_{\alpha \beta} = \frac{1}{2} \eta^{\beta\gamma}  \partial_{\alpha} A_{\beta \gamma}$, the Einstein field equations with the first order of the Ricci tensor read
\begin{equation} \label{eq:ricci1}    
R_{\alpha \beta}^{(1)}  \approx - \frac{1}{2}\partial_{\gamma}\partial^{\gamma} \Phi_{\alpha \beta} \approx 8\pi \mathrm{G} \left(P_{\alpha \beta}  -  \frac{P}{2}g_{\alpha\beta}\right),
\end{equation} 
where $\mathrm{G}$ is the Newtonian gravitational constant and $P_{\alpha\beta}$ is the stress-energy tensor. For a point particle `\textit{i}' located at $x_i = (t, \vec{r}_i)$ with energy $m_i$ and 4-velocity $u^{\alpha} := dx_i^{\alpha}/d\tau$ (with respect to the proper time $d\tau$), the 4-momentum is $P^{\alpha} := m_iu^{\alpha}$. If we consider a perfect fluid with $N$ particles located at $x_i \approx x_0 = (t, \vec{r}_0 \equiv 0)$ for all, with total mass $m = \sum_i m_i$ and moving in parallel with velocity $u^\alpha$, the density $\rho(x)$ and the stress-energy $P^{\alpha\beta}(x)$ are \cite{Brown1993}: 
\begin{equation}\label{eq:stress}
\begin{split}
\rho(x) & := \sum_{i=1}^{N} \int \frac{\delta^4(x-x_i(\tau))}{\sqrt{|g(x)|}}m_i\, d\tau \; \approx \;   \Theta (t-r/c)\,\delta^3\left(\vec{r}\right)\, m
\\
P^{\alpha\beta}(x) & = \hat{\rho}(x) u^{\alpha} u^{\beta} - \sigma(x) g^{\alpha\beta}  = {\rho}(x)((1+w)u^{\alpha} u^{\beta} - w g^{\alpha\beta}), 
\end{split}
\end{equation}
where $x = (t,\vec{r})$ are the measurement coordinates with $r := ||\vec{r}|| > 0$, $\delta^n$ is the \textit{n}-dimensional Dirac delta, $\Theta (t-r/c) \approx \Theta(t) :=\int _{-\infty}^{t}\delta (s)\,ds$ is the Heaviside step function by neglecting the retarded term $r/c \ll t$ with $c \equiv 1$, $w := \sigma/\rho$ is the equation of state, and $\hat{\rho} := \rho(x) + \sigma(x) = \rho(x)(1+w)$ is the \textit{total energy density}, which consists of the \textit{rest energy density}, $\rho(x)$, and the \textit{pressure}, $\sigma(x) := n\partial\rho(x)/\partial n - \rho(x)$, being $n := \mathrm{d}\rho/\mathrm{d}V$ the particle number density. Since $P = \rho[(1+w)-4w] = \rho(1-3w)$, the matter tensor is\begin{equation} 
    {\mathfrak P}_{\alpha\beta} := P_{\alpha\beta} - \frac{P}{2} g_{\alpha\beta} \approx (1+w)\rho(x) \underbrace{\left(u_{\alpha}u_{\beta}  -  \frac{1}{2}\left(\frac{1-w}{1+w}\right)\eta_{\alpha\beta}\right)}_{U_{\alpha\beta}{(w)}}\,,
\end{equation} with $w \ge 0$ and $U_{\alpha\beta}{(w)} := u_{\alpha}u_{\beta}  -  \frac{1}{2}\left(\frac{1-w}{1+w}\right)\eta_{\alpha\beta}$. Thus, the solution of the first-order Einstein equations (Eq. \ref{eq:ricci1}) is given by the Green function $\Phi_{\alpha \beta}$ with d'Alembert operator $\partial_{\gamma}\partial^\gamma$ whose application is proportional to $\delta^3\left(\vec{r}\right)$ in a similar way to the Lienard-Wiechert fields, that is,
\begin{eqnarray}  \label{eq:einstein1.sol1} 
   - \partial_{\gamma}\partial^\gamma  \Phi_{\alpha \beta} \left({\vec {r}},t\right) \; &  \approx &\; 16\pi\mathrm{G}\, \Theta (t)\,\delta^3\left(\vec{r}\right)\, \hat{m}\, U_{\alpha\beta}(w) 
   \\ \nonumber
     \Phi_{\alpha \beta}\left({\vec {r}},t\right) & \approx & - 16\pi\mathrm{G} { \frac{1}{4\pi r}} \,\hat{m}\, U_{\alpha\beta}(w) + K_{\alpha\beta} \;\approx \; 4\phi(r) U_{\alpha\beta}(w) + K_{\alpha\beta},
\end{eqnarray}
where $\phi(r) := -\mathrm{G}\hat{m}/r$ is the static potential by neglecting the Lienard-Wiechert retarded term, $\hat{m} := (1+w)m$, and $K_{\alpha\beta}$ are constants of integration. For purely static dust ($w=0$, $u_0 = 1$, $u_i = 0$ for $i \in \{1,2,3\}$), and by recalling that $\eta_{\alpha \beta} = \mathrm{diag}(1,-1,-1,-1)$, we have the following.
\begin{equation} \label{eq:einstein1.sol2}   
\Phi_{0 0} = \Phi_{i i} = 2\phi(r) + K = -\frac{2\mathrm{G{m}}}{r} + K,
\end{equation}
where $K := K_{00} = K_{ii}$. Therefore, the Newtonian gravitation field is recovered, and the perturbation $\Phi_{\alpha\beta}$. On the other hand, any double-copy perturbation requires that $w=1$, whence \begin{equation} \label{eq:metric_perturbation}
\Phi_{\alpha \beta} \approx 4\phi(r) u_{\alpha}u_{\beta} + K_{\alpha\beta}.
\end{equation}


\textit{Derivation from the Lagrangian formulation}. The classical Newtonian gravity can be obtained by using the extremal theory for the first order perturbation of the metric $g$. Let $m$ and $x^{\alpha}$ be the mass and the position, respectively, of a test point particle (with state $w_m$) living in a spacetime perturbed by a (source) mass $M$ (with state $w$). The action functional for this particle is
\begin{eqnarray}\label{eq:action0}  
S = \int m d\tau = \int m d\lambda \sqrt{g_{\mu\nu}\frac{dx^{\mu}}{d\lambda}\frac{dx^{\nu}}{d\lambda}}  \approxeq \int \frac{m}{2} g_{\mu\nu}u^{\mu}u^{\nu} d\tau,
\end{eqnarray} 
where $u^{\mu} := dx^{\nu}/d\lambda$ and the homomorphism between $\delta \sqrt{\bullet} = 0$ and ${\frac{1}{2}} \delta {\bullet} = 0$ has been used. Now, the gravitational field is $\phi := -\mathrm{G}\hat{M}/r < 0$, with $|g_{\alpha\alpha}| \le 1$ (classical spacetime gravity), and the Lagrangian functional is 
\begin{equation}\label{eq:lagrangian1b}
\begin{split}
 L & = {\frac{m }{2}} g_{\alpha \beta} u^\alpha u^\beta \approx {\frac{m}{2}} \left(\eta_{\alpha\beta}+4\phi(r)\left(u_{\alpha}u_{\beta}-\left(\frac{1-w}{1+w}\right)\frac{\eta_{\alpha\beta}}{2}\right)\right)u^{\alpha}u^{\beta}\\
 & ={\frac{m }{2}} \eta_{\alpha\beta}u^{\alpha}u^{\beta} + m \left(\frac{1+3w}{1+w}\right)\phi(r)={\frac{m}{2}}\eta_{\alpha\beta}u^{\alpha}u^{\beta} + m I_w \phi(r),
\end{split}
\end{equation}
where $I_w := (1+3w)/(1+w)$. The corresponding Euler--Lagrange equations are the geodesic equations, $du^{\mu}/d\tau = -\Gamma^{\mu}_{\alpha \beta} u^\alpha u^\beta$, but this approach leads to 
\begin{equation}\label{eq:Euler.lagrange0} 
\frac{d}{d\tau} \frac{\partial L}{\partial u^{\alpha}} - \frac{\partial L}{\partial x^{\alpha}} = 0
\hspace{4mm} \Rightarrow \hspace{4mm}  f := m \frac{du_{\alpha}}{d\tau} = m  I_w \frac{\partial \phi(r)}{\partial x^{\alpha}}.
\end{equation}
which corresponds to the classical Newtonian force at the limit $w=0$ (purely dust particles). By taking spherical coordinates with radius $r$, the classical gravitational energy is defined by the work,
\begin{equation}\label{eq:Euler.lagrange2} 
m '(r) := \int_{r_m}^r f(r') dr' =  m I_w \int_{r_m}^r \frac{\partial \phi(r')}{\partial r'} dr'= m I_w \Delta \phi(r),  \end{equation} where $r_m$ is the origin of the energy potential and $\Delta \phi(r) := \phi(r) - \phi(r_m)$. It is usual to choose $r_m \to \infty$ as the energy origin, but we purposely set $r_m = \mathrm{G}I_w m$ or $r_m =  \mathrm{G}I_w \hat M$, were $w$ depends on the source $M$. Therefore, the (semi-)classical gravitational-energy perturbation is\begin{equation} 
\label{eq:perturbed_energy}
m '(r) =  m I_w \left(-\frac{\mathrm{G}\hat{M}}{r} + \frac{\mathrm{G}\hat{M}}{r_m} \right).
\end{equation}

The total potential energy for a test point particle depends on the origin chosen. If $r_m = \mathrm{G}I_w \hat{M}$ is set as the source origin, then the semi-classical potential energy, $m'(r) = m + m\phi(r)$, approaches to $\lim_{r \to \infty}m'(r) = m$, i.e., the test particle self-energy. By changing the origin to the position of the test particle, $r_m = \mathrm{G}I_w {m}$, the total energy, $m'(r) = M + m\phi(r)$, represents the total source energy when $\lim_{r \to \infty}m'(r) = M$. Therefore, for both cases, the total contribution of gravity is zero. In any way, the kinetic and gravitational energies in GR depend on the reference system or the connection. For instance, in teleparallel gravities it is found that the total energy of coupled matter and gravity is proportional to the 3-space Ricci scalar up to a first-order perturbation \cite{Abedi2015}.

\subsection{Linearized gravitomagnetism}
\label{sec:Gravitomagnetism}

\textit{Separable perturbation}. Let $\Phi_{\alpha\beta}(t, r)$ be a two-particle metric perturbation that depends on time $t$ and on the relative position of the point particles, $r$. By assuming that the only dependence in $t$ is given by the 4-velocity, $u_{\alpha} = u_{\alpha}(t)$, the solution to the first-order Einstein field equations for perfect fluids is quasi-stationary, i.e., it is similar to the Eq. \ref{eq:einstein1.sol1}. This implies that, like the stress-energy source tensor $P_{\alpha\beta} = \hat{\rho}u_{\alpha}u_{\beta} - \frac{w}{1+w}\hat{\rho}\eta_{\alpha\beta}$ and the source tensor $P_{\alpha\beta} - \frac{1}{2} P\eta_{\alpha\beta} =  \hat{\rho}u_{\alpha}u_{\beta} - \frac{1-w}{2(1+w)}\hat{\rho}\eta_{\alpha\beta}$, the metric perturbation can be decomposed into two terms: a term $\bar{A}_{\alpha\beta}$ proportional to the 4-velocity tensor, $u_{\alpha}u_{\beta}$ (for instance, using a 4-potential, $\bar{A}_{\beta}:=\bar{A}u_{\beta}$ with static field $\bar{A} := 2\sqrt{-\phi}$), and another term $\bar{B}_{\alpha\beta}$ proportional to the flat metric tensor, $\eta_{\alpha\beta}$, and to another static field, $\bar{B}$ (see, for instance, Eq.~\ref{eq:einstein1.sol1}). Therefore,
\begin{eqnarray} \label{eq:metric.pert}  
\Phi_{\alpha\beta} \approx \bar{A}_{\alpha\beta} +  \bar{B}_{\alpha\beta}
, \hspace{5mm}
\bar{A}_{\alpha\beta} := \bar{A}_{\alpha}\bar{A}_{\beta} , \hspace{5mm}
\bar{B}_{\alpha\beta} := -\bar{B}^2\eta_{\alpha\beta},
\end{eqnarray}where one can identify the perturbation $\bar{A}_{\alpha\beta}$ as in the Kaluza--Kein metric, but in four dimensions, that is, as a double-copy linearized gravity \cite{Gurses2018}.

\vspace{5mm}

\noindent \textit{Maxwell-like equations}. The first order Einstein field equations (Eq. \ref{eq:ricci1}) for quasi-stationary perfect fluids can be split into two parts: 
\begin{eqnarray} \label{eq:einstein2b}  
\hspace{30mm}
\partial_{\gamma}\partial^{\gamma}\bar{A}_{\alpha\beta} \approx & -16\pi\mathrm{G}\hat{\rho}u_{\alpha}u_{\beta},
\\ \label{eq:einstein2c} 
\hspace{30mm}
\partial_{\gamma}\partial^{\gamma}\bar{B}_{\alpha\beta} \approx & 16\pi\mathrm{G}\frac{1-w}{2(w+1)}\hat{\rho}\eta_{\alpha\beta},
\end{eqnarray}
where the harmonic coordinates condition has been chosen. Taking $w=1$, Eq.~\ref{eq:einstein2b}  is very similar to the Maxwell equations in the Lorentz gauge when replacing the electric 4-current, $j^\mu := \hat{\rho} u^\mu$, by a tensor, $J^{\mu\nu} := \hat{\rho} u^\mu u^\nu$.

\vspace{5mm}

\noindent \textit{Lorentz-like force}. Similar dynamics to the Lorentz force can be obtained from the geodesic equations at the first order:
\begin{eqnarray} \label{eq:geodesic0}  
\frac{d u^{\gamma}}{d\tau} \approx - \Gamma^{\gamma}_{\alpha\beta}{}^{(1)}u^{\alpha}u^{\beta} \approx - ({\Gamma}^{\gamma}_{\alpha\beta}{}^{(A)}+{\Gamma}^{\gamma}_{\alpha\beta}{}^{(B)})u^{\alpha}u^{\beta},
\end{eqnarray}where $\Gamma^{\gamma}_{\alpha\beta}{}^{(A)}$ and $\Gamma^{\gamma}_{\alpha\beta}{}^{(B)}$ are the first order Christoffel symbols for the the metric perturbations $\bar{A}_{\alpha\beta}$ and $\bar{B}_{\alpha\beta}$, respectively.

By multiplying $u^{\alpha}u^{\beta}$ on both sides of the equations, and by recalling that $u_{\gamma}\partial_{\nu}u^{\gamma} = 0$ and $u^{\alpha}\partial_{\alpha}\bar{A} = u^{\alpha}\partial_{\alpha}\bar{B}  = 0$, where $\bar{A}$ and $\bar{B}$ are static fields, the following equation is obtained:
\begin{equation} \label{eq:geodesic.gem0}   
\frac{d u^{\gamma}}{d\tau} \approx  \bar{A}^{\mu} \bar{F}_{\mu}^{\ \gamma}  +  \bar{B}\partial^{\gamma}\bar{B},
\end{equation} where $\bar{F}_{\beta\nu} := \partial_{\beta}\bar{A}_{\nu} - \partial_{\nu}\bar{A}_{\beta}$. The case $\bar{B} = 0$ is equivalent to the Lorentz force law. The gravitational Lorentz force can be also obtained by using a Lagrangian functional with $B=0$, that is, $\phi = -\bar{A}^2/2 = -\eta_{\alpha\beta}\bar{A}^{\alpha}\bar{A}^{\beta}/2$. In this case, we have
\begin{equation} \label{eq:lagrangian1c}   
L \approx \frac{m}{2}g_{\alpha\beta}u^\alpha\beta \approx \frac{m}{2}(1-\bar{A}^2)\eta_{\alpha\beta}u^{\alpha}u^{\beta}.
\end{equation} The corresponding Euler--Lagrange equations (Eq. \ref{eq:Euler.lagrange0}) are
\begin{equation} \label{eq:Euler.lagrange1b}  
\begin{split}
\frac{d}{d\tau}\left(m u_{\alpha} - m\bar{A}\bar{A}_{\alpha} \right) + m\bar{A}^{\gamma}\frac{\partial  \bar{A}_{\gamma}}{\partial x^{\alpha}}=0,\\
 \frac{d}{d\tau}\left(m u_{\alpha}\right) - m\bar{A}^{\gamma} \frac{\partial A_{\alpha}}{\partial x^{\gamma}}  + m\bar{A}^{\gamma}\frac{\partial  \bar{A}_{\gamma}}{\partial x^{\alpha}} = \frac{d}{d\tau}\left(m u_{\alpha}\right) - m\bar{A}^{\gamma}\bar{F}_{\gamma\alpha}=0,
\end{split}
\end{equation}
where the conditions $d\bar{A}/d\tau = 0$ and $d/d\tau = u^{\gamma}\partial_{\gamma}$ have been used.

\section{Second-order of point Lorentz force}
\label{appendix:second-order-lorentz}

From the exact solution of the metric (Eq. \ref{eq:exactsolution}), the Lagrangian reads
\begin{equation} \label{eq:lagrangian3}   
L = \frac{m}{2}g_{\alpha\beta}u_2^{\alpha}u_2^{\beta}
\hspace{3mm} = \hspace{3mm}  \frac{m}{2}\eta_{\alpha\beta}u_2^{\alpha}u_2^{\beta} + A_1^{\gamma} q_2 u_{2\gamma} - \frac{4\pi\mathrm{G}}{\mathrm{k}} m A_1^{\gamma} A_{2\gamma}
\end{equation}
The corresponding Euler--Lagrange equations are
\begin{equation}\label{eq:Euler.lagrange3}
\begin{split}
 \frac{d}{d\tau}\left(m u_{2\alpha}\right) & + q_2u_2^{\gamma} \frac{\partial A_{1\alpha}}{\partial x_2^{\gamma}}  - q_2u_2^{\gamma}\frac{\partial  A_{1\gamma}}{\partial x_2^{\alpha}} +  \frac{4\pi\mathrm{G}}{\mathrm{k}} m \frac{\partial}{\partial x^\alpha_2}\left( A_1^{\gamma} A_{2\gamma}\right)\\
&= \frac{d}{d\tau}\left(m u_{2\alpha}\right) + q_2u_2^{\gamma}\mathcal{F}_{\gamma\alpha} +  \frac{4\pi\mathrm{G}}{\mathrm{k}} m \frac{\partial}{\partial x^\alpha_2}\left( A_1^{\gamma} A_{2\gamma}\right)=0,
\end{split}
\end{equation}
which correspond to the second-order correction of the Lorentz force. For instance, let $A_{2\gamma} = \mathrm{k}q_2/r$ be constant when the distance is $r = 1$m with respect to a proton ($m_p= 938 \mathrm{MeV} ,\, q = 1$e) where a Synchrotron is placed. Therefore
\begin{equation} \label{eq:Lorentz2}    
\Delta \left(m_p u_{2\alpha}\right) \;\;=\;\; - q_2u_2^{\gamma}\mathcal{F}_{\gamma\alpha} \,\Delta\tau +  4\pi\mathrm{G} \frac{m_p }{r \textcolor{black}{c^4}} q_2 u_{2\gamma}\frac{\partial A_1^{\gamma}}{\partial  x^\alpha_2}  \,\Delta\tau.
\end{equation} Notice that the light speed $\textcolor{black}{c}\equiv 1$ is highlighted only to conveniently remind the units, although natural units are chosen in this work. To estimate the order of magnitude of the correction, one can take for instance a proton accelerated by a Synchrotron up to $m_p\,u_{2\alpha} = 469 \mathrm{GeV} = 500 \,m_p$, that is, a relativistic 4-velocity of $u_{2\alpha} = \gamma v_{\alpha} = 500$, where $v_{\alpha} = 0.999998$ and $\gamma = 1/\sqrt{1-v_{\alpha}^2}$. Therefore, $- q_2 u_{2\gamma} \mathcal{F}_{\gamma\alpha} \,\Delta\tau \sim q_2 u_{2\gamma} {\partial A_1^{\gamma}}/{\partial  x^\alpha_2}  \,\Delta\tau \sim  500 \mathrm{GeV}$, and the final order of magnitude is\begin{equation}
    \Delta \left(m_p u_{2\alpha}\right) \;\;\sim\;\; - q_2u_2^{\gamma}\mathcal{F}_{\gamma\alpha} \,\Delta\tau\left( 1 -  4\pi\mathrm{G} \frac{m_p }{r \textcolor{black}{c^4}}  \right),
\end{equation} which is a correction by gravity absolutely negligible for the proton.

\end{appendices}


\end{document}